\documentclass[prx,superscriptaddress,aps, bibliography, twocolumn, reprint, showpacs, longbibliography, footnoteinbib]{revtex4-1}

\usepackage{enumerate}
\usepackage{braket}  
\usepackage{graphicx}
\usepackage{amssymb} 
\usepackage{amsfonts}
\usepackage{amsmath}
\usepackage{dsfont}
\usepackage{natbib}
\usepackage{dcolumn} 
\usepackage{bm}  
\usepackage[mathscr]{eucal}
\usepackage[dvipsnames]{xcolor}
\usepackage[colorlinks, linkcolor=OrangeRed,citecolor=RoyalBlue,urlcolor=NavyBlue]{hyperref}
\usepackage[all]{hypcap} 
\usepackage{mathtools}
\usepackage{wasysym}

\newcommand{\tTh}{t_{\text{Th}}}
\newcommand{\Xf}{X_{\text{front}}}
\newcommand{\N}{\mathcal{N}(x)}
\newcommand{\Pxt}{\Pi(x,t)}
\newcommand{\Pxinf}{\Pi(x,\infty)}
\newcommand{\PxO}{\Pi(x,0)}
\newcommand{\Rt}{\Pi(0,t)}
\newcommand{\Rinf}{\Pi(0,\infty)}
\newcommand{\WAT}{W_{\text{AT}}}

\newcommand{\be}{\begin{equation}}
\newcommand{\ee}{\end{equation}}
\def\ba{\begin{aligned}}
\def\ea{\end{aligned}}
\newcommand{\bea}{\begin{eqnarray}}
\newcommand{\eea}{\end{eqnarray}}
\def\bes{\begin{subequations}}
\def\ees{\end{subequations}}
\def\bal{\begin{align}}
\def\eal{\end{align}}

\newcommand{\lb}{\left[}
\newcommand{\rb}{\right]}

\newcommand{\rev}[1]{{\color{black}#1}}
\newcommand{\revB}[1]{{\color{black}#1}}

\usepackage{soul}
\usepackage{ulem} 

\begin{document}

\title{Subdiffusion in the Anderson model on random regular graph}
 \author{Giuseppe De Tomasi}
 \email[]{detomasi@pks.mpg.de}
 \affiliation{Department of Physics, T42, Technische Universit\"at M\"unchen,
 James-Franck-Stra{\ss}e 1, D-85748 Garching, Germany}
 \author{Soumya Bera}
 \affiliation{Department of Physics, Indian Institute of Technology Bombay, Mumbai 400076, India}
 \author{Antonello Scardicchio}
 \affiliation{Abdus Salam International Center for Theoretical Physics, Strada Costiera 11, 34151 Trieste, Italy}
 \affiliation{INFN, Sezione di Trieste, Via Valerio 2, 34126, Trieste, Italy}
 \author{Ivan M. Khaymovich}
 \affiliation{Max-Planck-Institut f\"ur Physik komplexer Systeme, N\"othnitzer Stra{\ss}e 38, 01187-Dresden, Germany}
\begin{abstract}
We study the finite-time dynamics of an initially localized wave-packet in the Anderson model on the random regular graph (RRG)
\rev{and show the presence of a subdiffusion phase coexisting both with ergodic and putative non-ergodic phase.
The full probability distribution $\Pxt$ of a particle to be at some distance $x$ from the initial state at time $t$,
is shown to spread} subdiffusively over a range of disorder strengths. 
\rev{The comparison of this result with the dynamics of the Anderson model on $\mathbb{Z}^{d}$ lattices, $d> 2$, which is subdiffusive only at the critical point implies that the limit $d\rightarrow \infty$ is highly singular in terms of the dynamics.}
A detailed analysis of the propagation of $\Pxt$ in space-time $(x,t)$ domain identifies four different regimes determined by the position of a wave-front $\Xf(t)$, which moves subdiffusively to the most distant sites $\Xf(t) \sim t^{\beta}$ with an exponent $\beta < 1$.
Importantly, the Anderson model on the RRG can be considered as proxy of the many-body localization transition (MBL) on the Fock space of a generic interacting system. In the final discussion, we outline possible implications of our findings for MBL.
\end{abstract}
\maketitle

%

{\it Introduction}---
The common belief that generic, isolated quantum systems thermalize as a result of their own dynamics has been challenged by a recent line of work\rev{s} showing that strong enough disorder can prevent them reaching thermal equilibrium~\cite{Basko06,gornyi2005interacting}.
This phenomenon, referred as  many-body localization (MBL)~\cite{Basko06,gornyi2005interacting, huse2015review, imbrie2017review, abanin2017recent,ALET2018498, CollAba}, generalizes the concept of Anderson localization~\cite{Anderson1958}
to the case of interacting particles, and has an important bearing on our understanding of quantum statistical mechanics.

Although MBL has been extensively studied~\cite{huse2015review, ALET2018498, CollAba}, many of its aspects are still under intense debate.
For example, only little is known on the nature of the MBL transition~\cite{Vosk15, altman2015review, Agar15, Vedi17, pietracaprina2017entanglement}. Recent numerical results show that the critical point of the transition may have been previously underestimated~\cite{Tik_MPS,vsuntajs2019quantum} and critical exponents extracted with exact numerics seems to violate general constraints (i.e.\ so-called Harris bounds)~\cite{huse2015review,chandran2015finite}.
%
%
Even the nature of the ergodic phase is not completely settled. For instance, subdiffusive dynamics has been observed on finite time scales and system sizes~\cite{Luitz16, Bar15, Pre17,Vipin2015, Scardi16, Gopa16, Khait16,schulz2019phenomenology}, but its mechanism and asymptotic limit are far from being clear~\cite{Bera17, Tik_MPS,gopalakrishnan2015low,Weiner19,detomasi2019dynamics}.
%

Numerically these difficulties originate from the exponentially increasing complexity of the problem with system sizes, which makes the resolution of these open issues an extremely hard task.
One way to overcome this problem is to consider approximate calculation methods like matrix product states~\cite{Tik_MPS,Khemani2015obtaining,Chandran2015spectral,YuMPS2015,LimMPS2015} in order to increase significantly system sizes.
Another way is to find proxies of interesting observables in more tractable models, which can reproduce the salient intrinsic features of MBL systems~\cite{Biroli2017Dynamics, Tikh2019Critical, Avetisov16, Logan19, Roy1, Roy2}.
%
\begin{figure}[t!]
\includegraphics[width=1.\columnwidth]{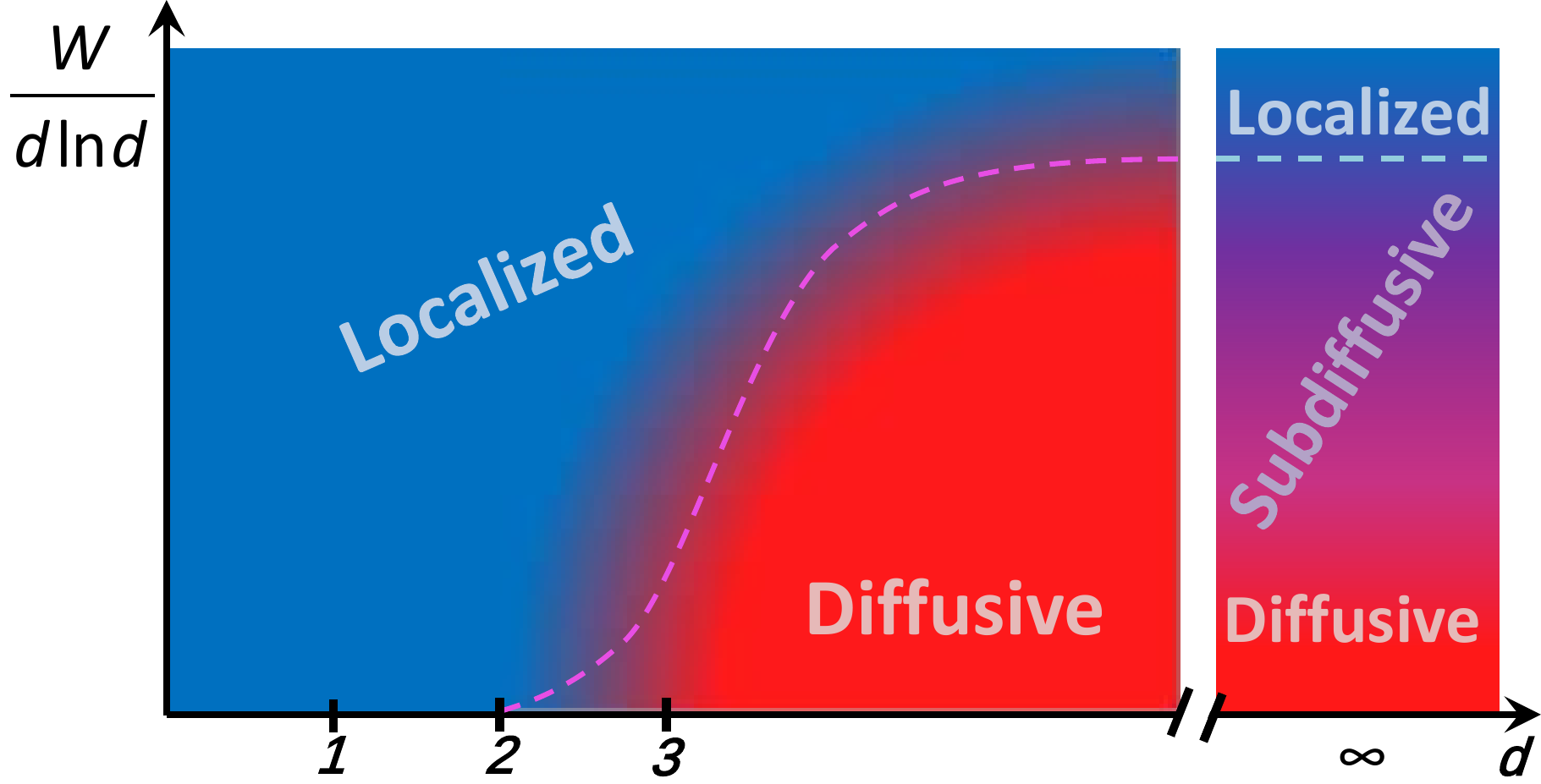}
\caption{\rev{{\bf \revB{Schematic} phase diagram of the anomalous transport of the Anderson model in $d$-dimension.} In $d=1,2$ the system is fully localized at any finite disorder. For $d>2$ the system has an Anderson transition at disorder strength $W=W_{AT}$, for $W<W_{AT}$ the transport is diffusive and subdiffusive only at the critical point.
\revB{At small $d-2\ll 1$ the critical disorder $W_{AT}\sim(d-2)$ (linear behavior of dashed line), while at large $d$ the it is given by $W_{AT} \sim d \ln d$ (dashed line saturation).} The limit $d\rightarrow \infty$ is given by the Anderson model on the RRG with the branching number $K$.
The latter limit is characterized by three distinct phases, a diffusive, subdiffusive and a localized one.
}}
\label{fig:Fig0}
\end{figure}
In this work we take the latter route considering
an Anderson model on hierarchical tree-like structure as a proxy for more realistic many-body systems.
Indeed, in \rev{\cite{Alt97
,Basko06}},  the idea of mapping interacting disordered electrons to an effective Anderson model on a section of Fock space was used to
give evidence of the stability of the MBL phase.
Recently, this paradigm and the hierarchical structure of the Fock space of generic many-body systems have revived interest in the Anderson problem on locally tree-like structures, such as the random regular graph (RRG).

Combining the hierarchical structure of the Fock space with the simplicity of regular graphs, i.e.\ the fixed branching number, RRG can be considered as a natural choice to approximate MBL systems~\cite{Biroli2017Dynamics, Tikh2019Critical, Avetisov16} and hope to overcome some of the numerical difficulties that have been mentioned earlier. Apart from the fact that RRG gives a new emphasis on the field of Anderson localization, independently also its own physics is extremely rich~\cite{Abou73,
Deluca14,Alt16,Alts16,Kra18,Biroli:2010wi,Lemarie17Small_K,Lemarie19two_loc_lengths,Parisi2018anderson,Savitz19Wegner_flow}. For instance, it has been shown that there is a possibility of a non-ergodic extended~(NEE) phase composed of critical states and placed between the ergodic and the localized phase~\cite{Deluca14, Alt16,Alts16,Kra18,Biroli2017Dynamics}.
Nevertheless, it has been argued that this NEE phase might merely be a finite-size effect and would disappear in thermodynamic limit~\cite{Biroli:2012vk,Biroli2018delocalization,Metz16PRL,Metz17PRB,
aizenman2011extended,Aizenman2013,Bapst2014HighConnectivity,
mirlin1991localization,Tik16,TikMir16,Sonner17,Tikh2019_K(w),Tikh2019Critical}.
However, this intricate question is far from being resolved.

Following the mapping of RRG to the Fock space of \rev{
many-body systems,
one expects the ergodic phase of wavefunctions on RRG to be} qualitative be mapped to the validity range of the eigenstate thermalization hypothesis (ETH)~\cite{Sred94,Deut91,Rigol2012} for many-body eigenstates.
Furthermore, it has been recently
suggested analytically and numerically confirmed that subleading corrections of ETH assumptions may lead to slow dynamics
of local observables after quench instead of diffusive one~\cite{Luitz16}.

Motivated by the above-mentioned mapping,  we study the spreading of an initially localized wave-packet in the Anderson model on the RRG as a probe of different dynamical phases.
In many-body systems, this can be considered as a proxy for the non-equilibrium dynamics of local operators after quench~\cite{Biroli2017Dynamics, monroe2015, Hauke15} and also as a direct measure for entanglement propagation~\cite{monroe2015, Hauke15}.
\rev{
We give
evidence of existence of subdiffusive dynamical phase over an entire range of parameters
both in a part of the phase diagram where most of the works
\cite{Alt16,Alts16,Kra18,Biroli:2010wi,Lemarie17Small_K,Lemarie19two_loc_lengths,Biroli:2012vk,Biroli2018delocalization,Metz16PRL,Metz17PRB,
aizenman2011extended,Aizenman2013,Bapst2014HighConnectivity,
mirlin1991localization,Tik16,TikMir16,Sonner17,Tikh2019_K(w),Tikh2019Critical}
agree
on ergodic nature of eigenstates
according to standard wave-function analysis
and in a putative non-ergodic phase~\cite{Deluca14, Alt16,Alts16,Kra18,Biroli2017Dynamics}.
%
Moreover, it is important to point out that the dynamics of the Anderson model on $\mathbb{Z}^{d}
$, $d>2$, is believed to be diffusive within its ergodic phase and subdiffusive only at the critical point~\cite{Ketzmerick1997,Ohtsuki1997}. Thus, the found subdiffusive phase in the limiting dimension $d\rightarrow \infty$ of the RRG provides a further example of the importance of dimensionality in the physics of localization, beside the well known example of fully localized systems in $d=1,2$, see Fig.~\ref{fig:Fig0}}.

{\it Model and methods}---
The Anderson model on the RRG is defined as
\be
\label{eq:Ham}
 \hat{H} = -\sum_{x,y\atop x \sim y }^{L} |x \rangle \langle y| + \sum_x^L h_x |x \rangle \langle x|,
\ee
where \rev{$x$ counts $L$ site states $|x\rangle
$ on the RRG.}
The first sum in $\hat{H}$ runs over sites $(x,y)$ that are connected ($x\sim y$) on the RRG with fixed branching  number (\rev{the number of neighbors of each site is fixed to $K+1=3$}). $\{h_x\}$ independent  random  variables distributed uniformly between  $[-W/2,W/2]$. This model is known to have an Anderson localization transition at $\WAT \approx 18.1 \pm 0.1
$~\cite{Kra18,Parisi2018anderson,Tikh2019Critical}.

We are interested in studying the full propagation of a wave function initially localized in a neighborhood of a site state $|x_0\rangle$, and having energy concentrated in a window of size $\delta E$ around the center of the band, $E=0
$.

\rev{A standard description for the dynamics employs} the distribution function $\Pxt$~\footnote{Notice the correct normalization $\sum_x \Pxt = 1$, and $\Pxt\ge0$.} which determines the probability to find the particle at time $t$ in some state at distance $x$ from the initial one
\be
\label{eq:Pi(x,t)_def}
 \Pxt = \overline{\frac{\sum_{y: d(y,x_0)=x} | \langle y | \hat{P}_{\Delta E} e^{-i \hat{H} t } \hat{P}_{\Delta E} | x_0 \rangle |^2}{\sum_{y} | \langle y | \hat{P}_{\Delta E} | x_0 \rangle |^2 }} \ .
\ee
The sum in Eq.~\eqref{eq:Pi(x,t)_def} runs over all states $|y\rangle
$ located at distance $d(y,x_0)=x$ from the initial state $|x_0\rangle$ . The distance $d(y,x_0)$ is defined as the shortest path's length that connects two sites on the RRG.
\rev{Importantly, in the many-body setting this distance is related to the Hamming metric of the Fock space~\cite{Alt97,ros2015integrals,monroe2015, Hauke15}.
The computation of the Hamming distance between two Fock states involves only the measure of local observables,  and has been measured experimentally in the MBL context, specially using it as a witness for entanglement propagation~\cite{monroe2015}.}

The overline in Eq.~\eqref{eq:Pi(x,t)_def} indicates the average over disorder, graph ensemble, and initial states $|x_0\rangle$.
$\hat{P}_{\Delta E} = \sum_{E\in \Delta E} | E\rangle \langle E|
$ is the projector onto eigenstates of $\hat{H}$ with energy $E$ \rev{from} a small energy shell
$E\in \Delta E = [-\delta E/2, \delta E/2]$ around the middle of the spectrum of $\hat{H}
$.
In particular, we consider $\delta E$ to be a small fraction $f$ ($f=1/8$) of the entire bandwidth
$E_{\textit{BW}}
$
($\delta E = f E_{\textit{BW}}
$).

The usage of the projector is motivated by several reasons.
First,
$\hat{P}_{\Delta E}$ avoids the
localized eigenstates at the edge of the spectrum \footnote{The single-particle mobility edges have been proven to exist at $W\gtrsim K$~\cite{aizenman2011extended}}.
Second, the initialization of the system in the microcanonical state with well-defined energy $E\in \Delta E$ in a small interval in the middle of the spectrum mimics ETH assumptions of many-body physics and under otherwise equal conditions prefers thermalization.
Thus, slow non-diffusive propagation of such projected wave-packet should rule out the possibility of fully ergodic phase (equivalent to random matrix theory~\cite{Mehta}).
Finally, the projector can be used as a dynamical indicator to distinguish a fully ergodic system from a non-ergodic one~\cite{bera19,gdt19}.
In a fully-ergodic phase, as a consequence of level repulsion, the return probability, $\Rt$, takes a standard form~\cite{gdt19} given by
\be
\label{eq:Return}
 \frac{\Rt}{\Pi(0,0)} = \left (\frac{\sin{\delta E t}}{\delta E  t}\right )^2.
\ee

The projector $\hat{P}_{\Delta E}$ slightly spreads the initial delta-function-like state $|x_0\rangle $
\rev{to the wave-packet $\hat{P}_{\Delta E}|x_0\rangle$ with a finite width.
}
This initialization supports the semiclassical description of wave-packet propagation in the system.
We ensure that our results do not change significantly with $\delta E$, provided it is not too big, please see Appendix~\ref{SM:Sec1_finite-size+f} for different $f$.

As a further measure of the spread of the wave-packet, we study the first moment of $\Pxt$,
\be
\label{eq:first_moment}
X(t) = \sum_x x \Pxt.
\ee
The \rev{wave-packet width at time $t=0$} induced by $\hat{P}_{\Delta E}$ can be simply estimated by
$X(0) = \sum_x x \PxO
$.

\begin{figure}
\includegraphics[width=1.\columnwidth]{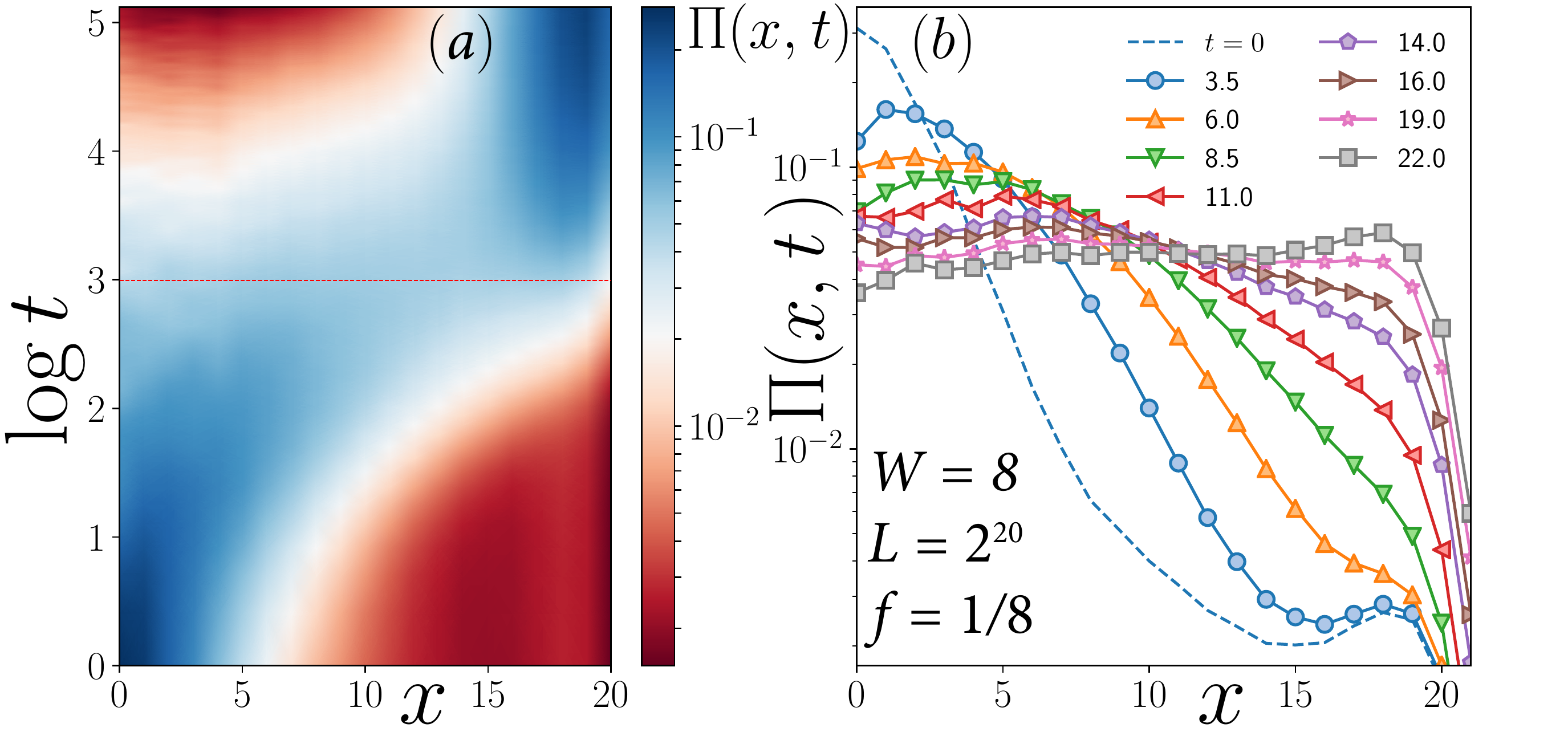}
\caption{{\bf Probability $\Pxt$ for the particle to be at distance $x$ from the initial state at time $t$.}
(a):~$\Pxt$ versus distance $x$ and time $t$ in a color plot.
Blue (red) color corresponds to high (low) values of $\Pxt$ and shows the propagation of the initially localized wave-packet with
the initial size $X(0) \simeq 2$
through the uniform distribution over distance $\Pi(x,t_{\textit{Th}})\simeq const$ (see the dashed line $t_{\textit{Th}} \approx 22$)
to the uniform distribution over sites $\Pxinf \simeq \N/L$.
(b):~Cross-section of the color plot in panel (a) at several times below $t_{\textit{Th}}$,
showing how the wave front propagates to the diameter of the graph.
All plots are shown at the most representative disorder amplitude $W=8$ for fixed system size $L=2^{20}$.
}
\label{fig:Fig1}
\end{figure}

\begin{figure}
\includegraphics[width=1.\columnwidth]{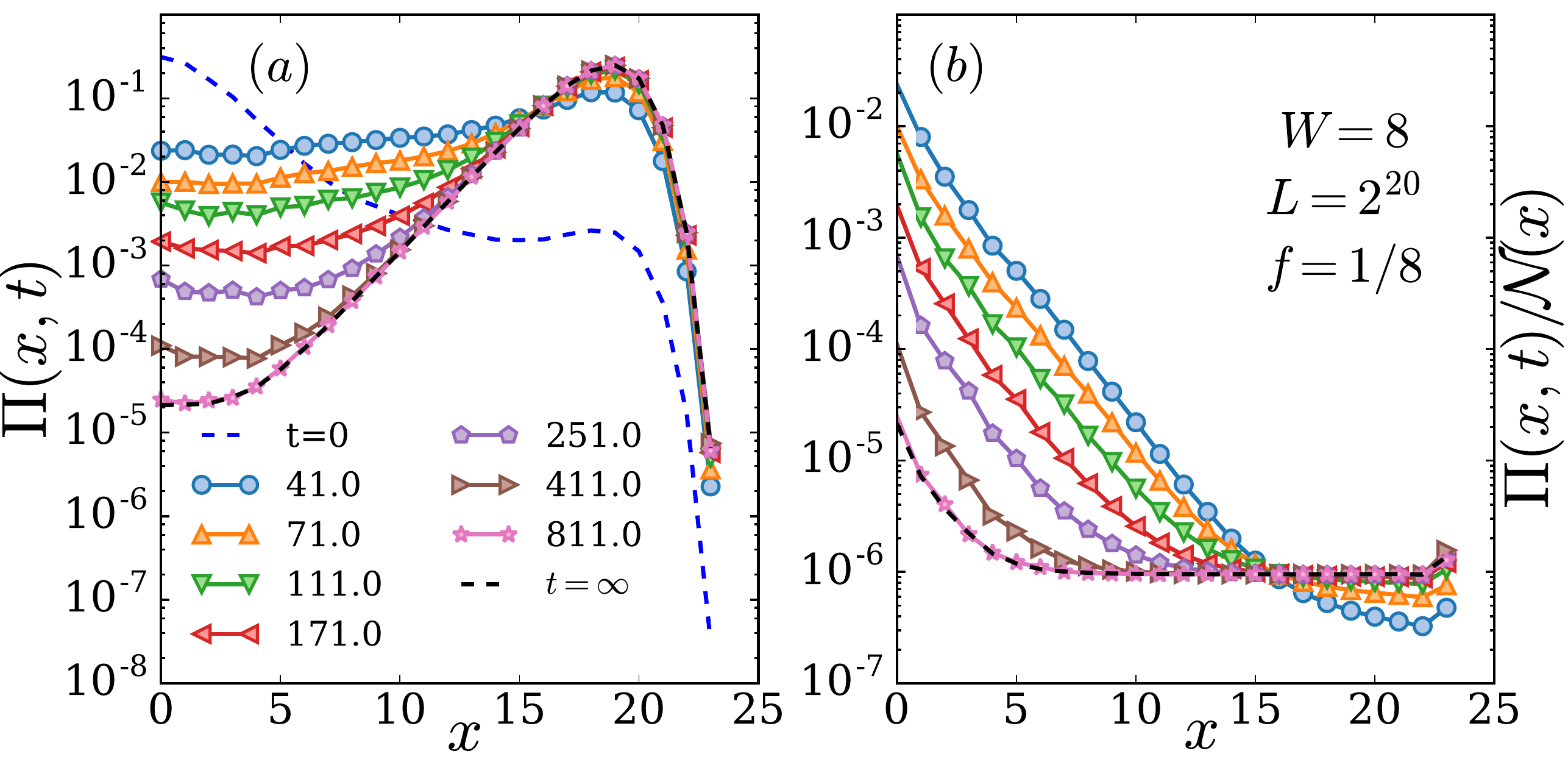}
\caption{
{\bf Probability $\Pxt$ versus distance $x$ at large times $t>t_{\textit{Th}}$.}
(a)~The cross-sections of Fig.~\ref{fig:Fig1}~(a) at several times above $t_{\textit{Th}}$, when the propagation front has already reached the diameter of the graph $\Xf> D=\ln{L}/\ln K$.
$\Pxt$ relaxes from the uniform distribution in the distance $\Pi(x,t_{\textit{Th}})\simeq const$
to the uniform distribution over sites $\Pxinf \simeq \N/L$.
Dashed line shows the initial distribution $\PxO$ as guide for eyes.
(b)~The distribution from panel (a) renormalized by the mean number of sites $\N$ at some distance $x$ from an initial site state $|x_0\rangle$.  This figure gives evidence of  the space-time factorization Eq.~\eqref{eq:Factorization}, once the front has already passed, $\Xf(t)>D$.
The parameters are the same as in Fig.~\ref{fig:Fig1}.
}
\label{fig:Fig2}
\end{figure}

\begin{figure}
\includegraphics[width=1.\columnwidth]{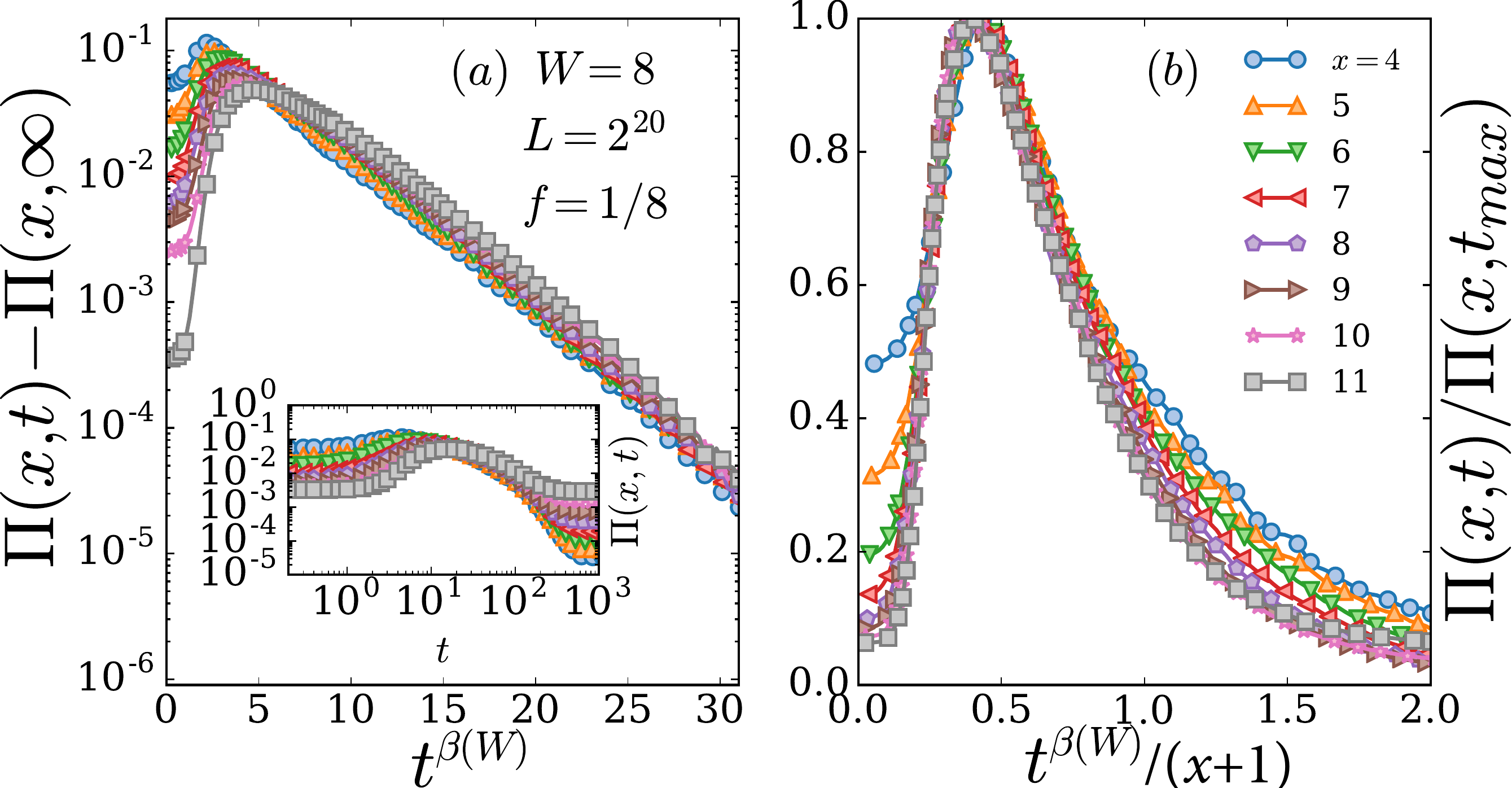}
\caption{{\bf Collapse of the probability $\Pxt$ versus $t$ at different distances $x$.}
(a)~Space-time factorization of $\Pxt$ Eq.~\eqref{eq:Factorization} 
at large times $t>t_{\textit{Th}}(x)$
corresponding to the relaxation inside the wave-packet $x<\Xf(t)$.
This relaxation is proportional to the return probability
$\Rt \sim e^{-\Gamma t^{\beta(W)}}$.
The time axis is properly rescaled with the power $\beta(W) = 1 - W/\WAT$, Eq.~\eqref{eq:beta},
to emphasize stretched exponential decay of $\Pxt$.
The inset shows $\Pxt$ on a log-log scale for all four stages of the evolution: initial distribution
$\Pxt \approx \PxO$ before wave front coming $x>\Xf(t)$, the maximum,
the common tail after front passing, and the eventual saturation.
(b)~Collapse of $\Pxt$ Eq.~\eqref{eq:front_X_f-x} around the wave-front confirming subdiffusive propagation $\Xf(t)\sim t^{\beta (W)}$.
The parameters are the same as in Fig.~\ref{fig:Fig1}.
}
\label{fig:Fig3}
\end{figure}

In Ref.~\cite{bera19} we have shown that for small values of $W$ ($0<W< 0.16 \WAT\simeq 3
$) the return probability $\Rt$ is consistent with the result of Eq.~\eqref{eq:Return}, confirming that the system is in a fully ergodic phase~\revB{\cite{FE-footnote, W3-footnote}}.
For larger disorder, $W\in [0.4 \WAT,0.7 \WAT] \simeq [8, 13]
$, $\Rt$ decays as a stretched exponential $\sim e^{-\Gamma t^{\beta(W)}}
$, where the exponent is well approximated by
\be\label{eq:beta}
\beta(W) \simeq 1-W/\WAT, \quad 0.4\WAT \lesssim W \lesssim 0.7 \WAT \
\ee
and goes to zero at the Anderson transition.
As a consequence, the drastic change in the time evolution of $\Rt
$ gives evidence of the existence of two \textit{dynamically distinct phases} (cf. Appendix~\ref{SM:Sec2_Weak_disorder} and~\ref{SM:Sec3_intermediate_W}).


As a side remark, before \rev{coming to the results, we 
stress the difference between the wave-packet propagation} on hierarchical structures and $d$-dimensional lattices like $\mathbb{Z}^d$.
It is well-known that
the return probability for a classical unbiased random walk on a Bethe lattice with branching number $K$ decays exponentially fast in time
$\sim e^{-\Omega(K) t}
$~\cite{Giacometti95} due to the exponential growth $K^x$ of the number of sites with the distance $x$ from an initial point $|x_0\rangle$.
Instead, in \rev{
$\mathbb{Z}^d$ lattices} the typical behavior is diffusive $\sim t^{-d/2}$ as the number of sites at distance $x$ grows
algebraically $\N \sim x^{d-1}
$.
Thus, the diffusive propagation on hierarchical tree lattices is characterized by a linear growth of the width of the wave-packet \ with time
$X(t)\sim t
$~(see Eq.~\eqref{eq:first_moment}) unlike $X(t)\sim t^{1/2}
$  in $d$-dimensional lattices.
Noticing this difference, we call the propagation in RRG subdiffusive if
$X(t)\sim t^{\beta}
$ with $\beta<1$.

In this work, we show that, as time increases, $\Pxt$ relaxes forming a wave-front $\Xf(t)$ that moves subdiffusively  to the most distant sites, as shown in Fig.~\ref{fig:Fig1}. More specifically the propagation of $\Pxt$ can be
divided into four regions
in space-time domain $(x,t)$ depending on the position of the moving front $\Xf(t)$:
\newline
(i)~At large distances (small times), $x>\Xf(t)
$, the wave-front has not yet \rev{reached} $x$, and
the distribution is nearly unperturbed
\be\label{eq:P(x,0)}
 \Pxt \approx \PxO,\qquad x>\Xf(t).
\ee
(see the red area at small times in Fig.~\ref{fig:Fig1} and the plateau at short times in the inset of Fig.~\ref{fig:Fig3}).
\newline
(ii)~At $x\simeq \Xf(t)
$ in proximity of the front propagation, $\Pi(x\lesssim \Xf(t),t)
$ renormalized by its maximal value $\Pi(\Xf(t),t)
$  collapses to an universal function
\be\label{eq:front_X_f-x}
{\Pxt-\Pxinf}= \Pi(\Xf(t),t) f(\Xf(t)-x) \,
\ee
with the semiclassical ($x,t$) front propagation governed by the parameter $\Xf(t)-x$, as shown in Fig.~\ref{fig:Fig3}~(b).

In particular, the front moves subdiffusively $\Xf(t) \sim t^{\beta(W)}
$, where $\beta(W)$ is given by Eq.~\eqref{eq:beta}~\footnote{For $W<0.16 \WAT$ the propagation is diffusive with $\beta=1$ and consistent with the fully-ergodic behavior given by Eq.~\eqref{eq:Return}.}.
\newline
(iii)~At larger times (smaller distances within the wave-packet), $x<\Xf(t)$,
$\Pxt$ shows space-time factorization
\be\label{eq:Factorization}
\Pxt-\Pxinf = g(x)\lb\Rt-\Rinf\rb,
\ee
with respect to the return probability $\Rt$ and a certain function $g(x)$, as shown in Figs.~\ref{fig:Fig2} and \ref{fig:Fig3}~(a). Thus in this regime, the relaxation is dictated by the return probability which is connected to the front of propagation by the following relation:
\be\label{eq:R(t)}
 \Rt\sim \exp\lb-\lambda\Xf(t)\rb,\qquad \lambda > 0.
\ee
(iv)~Eventually at very long times $\Pxt$ saturates at the uniform distribution over sites $\Pxinf=\N/L$, where
$\N\sim K^x$ is the mean number of sites at some distance $x$ from an initial site state $|x_0\rangle$ and $L$ is the number of sites, Figs.~\ref{fig:Fig2}.
\newline
Stages~(i) and~(ii) are presented only for times $t<t_{\textit{Th}}$ corresponding to front propagation inside the graph $\Xf(t)<D$, where $D \simeq \ln L/\ln K
$ is the diameter of the graph. At larger times only relaxation with the return probability~(iii) and saturation~(iv) stages are relevant.


\begin{figure}
\includegraphics[width=1.\columnwidth]{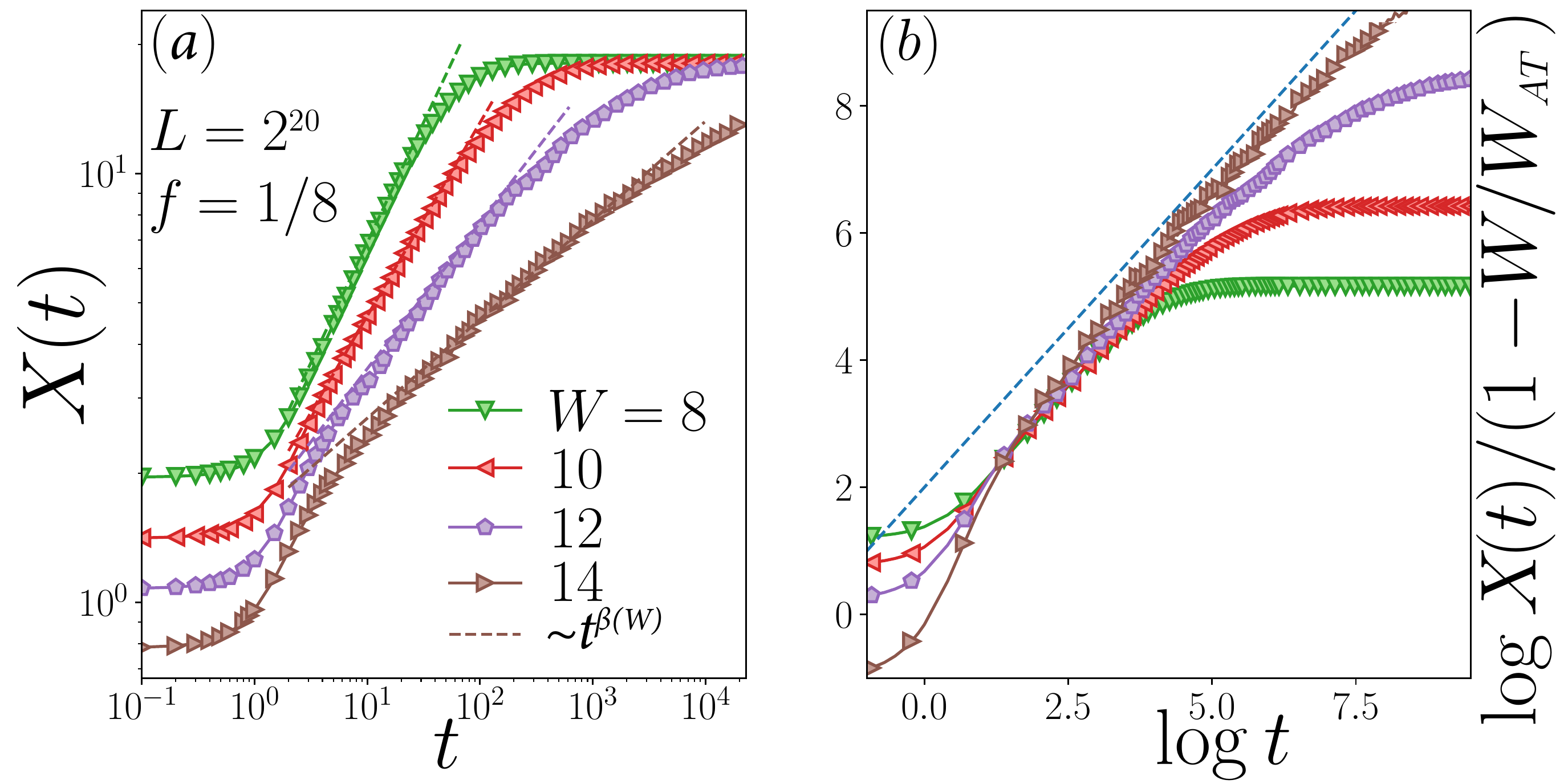}
\caption{{\bf Subdiffusive wave-packet spreading.}
(a)~Wave-packet width $X(t) = \sum_x x\Pxt$ versus time $t$ on a log-log scale for different disorder strengths supplemented by guide for eyes $\sim t^{\beta(W)}$ (dashed lines), with $\beta (W)=1-W/\WAT$, Eq.~\eqref{eq:beta}.
(b)~Collapse of wave-packet width $\ln{X(t)}/\beta (W)$ from panel (a)
showing the unit slope versus $t$ (dashed line) in increasing time interval with growing disorder amplitude $W$.
The parameters are the same as in Fig.~\ref{fig:Fig1}.
}
\label{fig:Fig5}
\end{figure}

{\it Results}---
We focus our attention on intermediate disorder values $8\leq W\leq 14
$,
for which the return probability shows slow dynamics 
$\Rt\sim e^{-\Gamma t^{\beta(W)}}
$, with $\beta(W)$ given by Eq.~\eqref{eq:beta} (see Appendix~\ref{SM:Sec3_intermediate_W}. 
\rev{
The propagation of $\Pxt$ versus distance $x$ and time $t$ at fixed system size $L=2^{20}\approx 10^6
$ is shown in Fig.~\ref{fig:Fig1}~(a).
At small times, $\delta E\ t\sim \mathcal{O}(1)
$, the wave-packet width is 
small} $X(0)\sim 2$, Eq.~\eqref{eq:first_moment}.
As time evolves, the wave-packet spreads in the form of wave-front
$\Xf(t)$~\footnote{$\Xf(t)$ it is given by the value of $x$ for which $\Pxt$ has a maximum} which transfers most of its weight to the most distant sites $(X(\infty)\simeq D
$) 
as shown in Fig~\ref{fig:Fig1}.

Although the dynamics on the RRG is typically not isotropic, the time scale $t_{\textit{Th}}$ at which the wave-front reaches the diameter could be seen as a natural choice for the Thouless time analogous to the time that a charge needs
to propagate through a diffusive conductor~\cite{Edwards_1972}.
The wave-front propagation at times $t<t_{\textit{Th}}$, $\Xf(t)<D),
$ can be seen in Fig.~\ref{fig:Fig1}~(a) and is explicitly shown in  Fig.~\ref{fig:Fig1}~(b).
Already for time $t_{\textit{Th}} \approx 22
$, as emphasized in the color plot of Fig.~\ref{fig:Fig1}~(a) (red dashed line), the main core has lost most of its amplitude $\Pi(x\le 5, t_{\textit{Th}})/\Pi(x\le 5, 0)\sim 10^{-1}
$
and \rev{$\Pxt$ becomes 
nearly uniform over the distance, $\Pi(x,t_{\textit{Th}})\simeq const
$.}

Figure~\ref{fig:Fig2}~(a) shows $\Pxt$ as function of $x$ at large times, $t>t_{\textit{Th}}
$, when the front has already reached the diameter of the graph.
In this regime, $\Pxt$ relaxes uniformly in distance $x$  to
the equiprobable configuration on the graph $\Pxinf =\frac{\N}{L}
$ (dashed line in Fig.~\ref{fig:Fig2}~(a)).
Thus, at these times $\Pxt-\Pxinf$ is factorized in $(x,t)$ \rev{
%
%
according to Eq.~\eqref{eq:Factorization}, with $g(0) = 1$ due to the uniform relaxation seen as well for $x=0$.}

\rev{
Detailed analysis shows that the factorization works beyond the limit}, $t>t_{\textit{Th}}
$, provided the wave-front crossed the observation point $x<\Xf(t)$, Fig.~\ref{fig:Fig3}~(a).
Subtracting from $\Pxt$ its long time limit $\Pxinf$, results in the collapse in Eq.~\eqref{eq:Factorization} of the curves for any $x<\Xf(t)$.
It is important to note that in Fig.~\ref{fig:Fig3} \rev{the time axis is rescaled as $t^{\beta (W)}$, with $\beta(W)$ given by Eq.~\eqref{eq:beta} to emphasize the stretched-exponential time relaxation shown to be true for the return probability in Ref.~\cite{bera19}, $\Rt-\Rinf \simeq e^{-\Gamma t^{\beta (W)}}
$.
Moreover, raw $\Pxt$ is shown in the inset of Fig.~\ref{fig:Fig3}~(a) on a log-log scale 
to demonstrate nearly unperturbed short-time behavior of $\Pxt$ in Eq.~\eqref{eq:P(x,0)}.
}

In order to analyze the time dependence of the wave-front propagation $\Xf(t)$ in Fig.~\ref{fig:Fig3}~(b) we collapse the curves dividing  $\Pxt$  by its maximum, $\Pi(\Xf(t),t)$, and rescale the time $t^{\beta (W)}$ \rev{in order to collapse the position of the maximum}. This collapse allows us to extract the following subdiffusive wave-front evolution
\be
\Xf(t) \simeq \Gamma(W)t^{\beta(W)},\qquad \beta(W)<1.
\label{eq:Front}
\ee
Moreover, the collapse of the curves, Fig.~\ref{fig:Fig3}~(b), implies the simple exponential dependence, Eq.~\eqref{eq:R(t)} 
of the return probability versus $\Xf(t)$ with a certain decay rate $\lambda$, $f(z) = e^{-\lambda z}
$, Eq.~\eqref{eq:front_X_f-x}~\footnote{Note that both collapses, Eqs.~\eqref{eq:front_X_f-x} and \eqref{eq:Factorization}, are consistent to each other
provided the exponential behavior of $f(z)$ in Eq.~\eqref{eq:front_X_f-x}, with the maximal value $\Pi(\Xf(t),t)$ given
by $g(\Xf)\lb \Rt - \Rinf \rb
$ due to the normalization condition of $\Pxt$ (see Appendix~\ref{SM:Sec4_Collapse} for more details).}.
\rev{
The front-propagation collapse, Eq.~\eqref{eq:front_X_f-x}, is shown to work also} for different disorder strengths in the range of interest $8\leq W\leq 13
$ (see Appendix~\ref{SM:Sec4_Collapse}).

As a further consequence,
the Thouless time, defined as the time when the wave-front reaches the graph diameter, scales as $t_{\textit{Th}} \sim (\frac{\ln{L}}{K})^{1/\beta(W)}
$.
The similar scaling of the Thouless time calculated for MBL systems in the subdiffusive phase~\cite{Serb_th17,Gar_th} supports the idea that wavepacket dynamics on RRG is a good proxy for MBL systems.

Finally,  we analyze the first moment $X(t)$, Eq.~\eqref{eq:first_moment}, of the radial probability distribution $\Pxt$.
Figure~\ref{fig:Fig5}~(a) shows  the algebraic growth of $X(t)$ in time for several $W$
\be\label{eq:first_moment_beta}
X(t)\sim t^{\beta (W)} \ ,
\ee
with the same subdiffusive exponent $\beta(W)$, Eq.~\eqref{eq:beta}, as in the wave-front propagation $\Xf(t)$.
Furthermore, the curves $X(t)$ can be reasonably well collapsed for the range of disorder strengths by considering the rescaled function $\frac{\ln{X(t)}}{\beta(W)}
$ versus $\ln t$.
\revB{This result is checked to be robust with respect to the finite-size effects and to the variation of the fraction $f$, see Appendix~\ref{SM:Sec1_finite-size+f}.}

\textit{Conclusion and Discussions}--
In this work, we
\rev{provide evidence of the existence of a subdiffusive phase for a finite range of parameters
by probing the dynamics of an initially localized particle on the RRG via} the probability distribution $\Pxt$ to detect it at distance $x$ at time $t$.

The relaxation of $\Pxt$ is characterized by the formation of a semi-classical wave-front $\Xf(t)$, \rev{moving} subdiffusively to the most distant sites.
\rev{
Remarkably,
as soon as the wave-front passed the observation point $x$,
the space-time factorization, $\Pxt=g(x)h(t)$, is found.
}

\rev{
The Anderson model on RRG gives the first example of an entire subdiffusive phase as
the systems on $\mathbb{Z}^{d}$ lattices are either localized for any finite disorder in $d=1,2$ or show subdiffusion only at the critical point, $d>2$~\cite{Ketzmerick1997,Ohtsuki1997}.}

It is important to note that the existence of a subdiffusive phase
is not in contrast with the possibility that the eigenfunctions are ergodic
in terms of the inverse participation ratio (IPR) scaling as the inverse of the volume.
To emphasize further, the IPR scaling is a statement about the nature of the fluctuations
of the eigenfunctions equivalent to the long time limit ($t \to \infty$) of
certain dynamical observable. On the other hand, in our study we probe the time evolution of a wavepacket and far away from the
aforementioned limit.
Thus, our study excludes the scenario that
the system is fully-ergodic at $W\ge 8$, which is a stronger requirement than just IPR ergodicity discussed above~\footnote{
\rev{In the literature~\cite{BogomolnyPLRBM2018,Nosov2019correlation,nosov2019mixtures,Baecker2019} this kind of phase is usually called {\it weakly} ergodic when the eigenfunctions occupy a finite, but small fraction $f_I\ll 1$ of the Hilbert space, meaning that the system splits into a finite number $\sim 1/f_I$ of nearly independent sectors which are ergodic within themselves, but not between each other.}}.

We have to mention that some works~\cite{Aizenman2012absolutely,Tikh2019_K(w)} claim only diffusive propagation ($\beta(W)=1$) for all $W<\WAT$. We do not report any crossover to diffusivity for our available system sizes and time scales.
Although we cannot completely rule out this possibility in the thermodynamic limit $L\to\infty$ and $t\to\infty$, the above finite-time subdiffusive dynamics is highly relevant for corresponding experiments in many-body systems.

In addition, the Anderson model on the RRG can be considered as a proxy for the dynamics of more realistic MBL systems. Our finding thus opens the possibility to have a subdiffusive dynamical phase in Fock space, that might imply slow relaxation of local observables.
\rev{This possible implication of subdiffusive spatial dynamics in MBL systems~\cite{Luitz16, Bar15, Pre17,Vipin2015, Scardi16, Gopa16, Bera17}
from slow Fock space dynamics
may give rise to a different mechanism which does not invoke the existence of Griffiths effects~\cite{Agar15, Gopa16, Vedi17}. 
Recent works (see, e.g.,~\cite{luitz2019multifractality}) show that in MBL systems the subdiffusive phase is also consistent with weakly-ergodic phase confirming RRG as commonly believed proxy.}


\begin{acknowledgments}
  We thank V.~E.~Kravtsov, F.~Pollmann, and S.~Warzel for helpful and stimulating discussions
  and M.~Haque and P.~A.~McClarty for careful reading of the manuscript.
  S.~B acknowledges support from DST, India, through Ramanujan Fellowship
  Grant No. SB/S2/RJN-128/2016 and ECR/2018/000876; S.~B. and G.~D.~T. also thank the
  visitor program of the Max Planck Institute for the Physics of Complex Systems, Dresden
  for hospitality during the final stage of the work.
  I.~M.~K. acknowledges the support of German Research Foundation
  (DFG) Grant No. KH~425/1-1 and the Russian Foundation for Basic Research Grant No. 17-52-12044.
\end{acknowledgments}

\bibliography{RRG_bib}

\begin{thebibliography}{90}%
\makeatletter
\providecommand \@ifxundefined [1]{%
 \@ifx{#1\undefined}
}%
\providecommand \@ifnum [1]{%
 \ifnum #1\expandafter \@firstoftwo
 \else \expandafter \@secondoftwo
 \fi
}%
\providecommand \@ifx [1]{%
 \ifx #1\expandafter \@firstoftwo
 \else \expandafter \@secondoftwo
 \fi
}%
\providecommand \natexlab [1]{#1}%
\providecommand \enquote  [1]{``#1''}%
\providecommand \bibnamefont  [1]{#1}%
\providecommand \bibfnamefont [1]{#1}%
\providecommand \citenamefont [1]{#1}%
\providecommand \href@noop [0]{\@secondoftwo}%
\providecommand \href [0]{\begingroup \@sanitize@url \@href}%
\providecommand \@href[1]{\@@startlink{#1}\@@href}%
\providecommand \@@href[1]{\endgroup#1\@@endlink}%
\providecommand \@sanitize@url [0]{\catcode `\\12\catcode `\$12\catcode
  `\&12\catcode `\#12\catcode `\^12\catcode `\_12\catcode `\%12\relax}%
\providecommand \@@startlink[1]{}%
\providecommand \@@endlink[0]{}%
\providecommand \url  [0]{\begingroup\@sanitize@url \@url }%
\providecommand \@url [1]{\endgroup\@href {#1}{\urlprefix }}%
\providecommand \urlprefix  [0]{URL }%
\providecommand \Eprint [0]{\href }%
\providecommand \doibase [0]{http://dx.doi.org/}%
\providecommand \selectlanguage [0]{\@gobble}%
\providecommand \bibinfo  [0]{\@secondoftwo}%
\providecommand \bibfield  [0]{\@secondoftwo}%
\providecommand \translation [1]{[#1]}%
\providecommand \BibitemOpen [0]{}%
\providecommand \bibitemStop [0]{}%
\providecommand \bibitemNoStop [0]{.\EOS\space}%
\providecommand \EOS [0]{\spacefactor3000\relax}%
\providecommand \BibitemShut  [1]{\csname bibitem#1\endcsname}%
\let\auto@bib@innerbib\@empty
\bibitem [{\citenamefont {Basko}\ \emph {et~al.}(2006)\citenamefont {Basko},
  \citenamefont {Aleiner},\ and\ \citenamefont {Altshuler}}]{Basko06}%
  \BibitemOpen
  \bibfield  {author} {\bibinfo {author} {\bibfnamefont {D.M.}\ \bibnamefont
  {Basko}}, \bibinfo {author} {\bibfnamefont {I.L.}\ \bibnamefont {Aleiner}}, \
  and\ \bibinfo {author} {\bibfnamefont {B.L.}\ \bibnamefont {Altshuler}},\
  }\bibfield  {title} {\enquote {\bibinfo {title} {Metal-insulator transition
  in a weakly interacting many-electron system with localized single-particle
  states},}\ }\href {\doibase https://doi.org/10.1016/j.aop.2005.11.014}
  {\bibfield  {journal} {\bibinfo  {journal} {Annals of Physics}\ }\textbf
  {\bibinfo {volume} {321}},\ \bibinfo {pages} {1126 -- 1205} (\bibinfo {year}
  {2006})}\BibitemShut {NoStop}%
\bibitem [{\citenamefont {Gornyi}\ \emph {et~al.}(2005)\citenamefont {Gornyi},
  \citenamefont {Mirlin},\ and\ \citenamefont
  {Polyakov}}]{gornyi2005interacting}%
  \BibitemOpen
  \bibfield  {author} {\bibinfo {author} {\bibfnamefont {I.~V.}\ \bibnamefont
  {Gornyi}}, \bibinfo {author} {\bibfnamefont {A.~D.}\ \bibnamefont {Mirlin}},
  \ and\ \bibinfo {author} {\bibfnamefont {D.~G.}\ \bibnamefont {Polyakov}},\
  }\bibfield  {title} {\enquote {\bibinfo {title} {Interacting electrons in
  disordered wires: Anderson localization and low-$t$ transport},}\ }\href
  {\doibase 10.1103/PhysRevLett.95.206603} {\bibfield  {journal} {\bibinfo
  {journal} {Phys. Rev. Lett.}\ }\textbf {\bibinfo {volume} {95}},\ \bibinfo
  {pages} {206603} (\bibinfo {year} {2005})}\BibitemShut {NoStop}%
\bibitem [{\citenamefont {Nandkishore}\ and\ \citenamefont
  {Huse}(2015)}]{huse2015review}%
  \BibitemOpen
  \bibfield  {author} {\bibinfo {author} {\bibfnamefont {R.}~\bibnamefont
  {Nandkishore}}\ and\ \bibinfo {author} {\bibfnamefont {D.~A.}\ \bibnamefont
  {Huse}},\ }\bibfield  {title} {\enquote {\bibinfo {title} {{Many-Body
  Localization and Thermalization in Quantum Statistical Mechanics}},}\
  }\href@noop {} {\bibfield  {journal} {\bibinfo  {journal} {Annual Review of
  Condensed Matter Physics}\ }\textbf {\bibinfo {volume} {6}},\ \bibinfo
  {pages} {15--38} (\bibinfo {year} {2015})}\BibitemShut {NoStop}%
\bibitem [{\citenamefont {Imbrie}\ \emph {et~al.}(2017)\citenamefont {Imbrie},
  \citenamefont {Ros},\ and\ \citenamefont {Scardicchio}}]{imbrie2017review}%
  \BibitemOpen
  \bibfield  {author} {\bibinfo {author} {\bibfnamefont {J.~Z}\ \bibnamefont
  {Imbrie}}, \bibinfo {author} {\bibfnamefont {V.}~\bibnamefont {Ros}}, \ and\
  \bibinfo {author} {\bibfnamefont {A.}~\bibnamefont {Scardicchio}},\
  }\bibfield  {title} {\enquote {\bibinfo {title} {Local integrals of motion in
  many-body localized systems},}\ }\href
  {http://onlinelibrary.wiley.com/doi/10.1002/andp.201600278/epdf} {\bibfield
  {journal} {\bibinfo  {journal} {Annalen der Physik}\ }\textbf {\bibinfo
  {volume} {529}},\ \bibinfo {pages} {1600278} (\bibinfo {year}
  {2017})}\BibitemShut {NoStop}%
\bibitem [{\citenamefont {Abanin}\ and\ \citenamefont
  {Papi{\'c}}(2017)}]{abanin2017recent}%
  \BibitemOpen
  \bibfield  {author} {\bibinfo {author} {\bibfnamefont {D.~A}\ \bibnamefont
  {Abanin}}\ and\ \bibinfo {author} {\bibfnamefont {Z.}~\bibnamefont
  {Papi{\'c}}},\ }\bibfield  {title} {\enquote {\bibinfo {title} {Recent
  progress in many-body localization},}\ }\href
  {https://doi.org/10.1002/andp.201700169} {\bibfield  {journal} {\bibinfo
  {journal} {Annalen der Physik}\ }\textbf {\bibinfo {volume} {529}},\ \bibinfo
  {pages} {1700169} (\bibinfo {year} {2017})}\BibitemShut {NoStop}%
\bibitem [{\citenamefont {Alet}\ and\ \citenamefont
  {Laflorencie}(2018)}]{ALET2018498}%
  \BibitemOpen
  \bibfield  {author} {\bibinfo {author} {\bibfnamefont {F.}~\bibnamefont
  {Alet}}\ and\ \bibinfo {author} {\bibfnamefont {N.}~\bibnamefont
  {Laflorencie}},\ }\bibfield  {title} {\enquote {\bibinfo {title} {Many-body
  localization: An introduction and selected topics},}\ }\href {\doibase
  https://doi.org/10.1016/j.crhy.2018.03.003} {\bibfield  {journal} {\bibinfo
  {journal} {Comptes Rendus Physique}\ }\textbf {\bibinfo {volume} {19}},\
  \bibinfo {pages} {498 -- 525} (\bibinfo {year} {2018})},\ \bibinfo {note}
  {quantum simulation / Simulation quantique}\BibitemShut {NoStop}%
\bibitem [{\citenamefont {Abanin}\ \emph {et~al.}(2019)\citenamefont {Abanin},
  \citenamefont {Altman}, \citenamefont {Bloch},\ and\ \citenamefont
  {Serbyn}}]{CollAba}%
  \BibitemOpen
  \bibfield  {author} {\bibinfo {author} {\bibfnamefont {D.~A.}\ \bibnamefont
  {Abanin}}, \bibinfo {author} {\bibfnamefont {E.}~\bibnamefont {Altman}},
  \bibinfo {author} {\bibfnamefont {I.}~\bibnamefont {Bloch}}, \ and\ \bibinfo
  {author} {\bibfnamefont {M.}~\bibnamefont {Serbyn}},\ }\bibfield  {title}
  {\enquote {\bibinfo {title} {Colloquium: Many-body localization,
  thermalization, and entanglement},}\ }\href {\doibase
  10.1103/RevModPhys.91.021001} {\bibfield  {journal} {\bibinfo  {journal}
  {Rev. Mod. Phys.}\ }\textbf {\bibinfo {volume} {91}},\ \bibinfo {pages}
  {021001} (\bibinfo {year} {2019})}\BibitemShut {NoStop}%
\bibitem [{\citenamefont {Anderson}(1958)}]{Anderson1958}%
  \BibitemOpen
  \bibfield  {author} {\bibinfo {author} {\bibfnamefont {P.}~\bibnamefont
  {Anderson}},\ }\bibfield  {title} {\enquote {\bibinfo {title} {{Absence of
  diffusion in certain random lattices}},}\ }\href {\doibase
  10.1103/PhysRev.109.1492} {\bibfield  {journal} {\bibinfo  {journal} {Phys.
  Rev.}\ }\textbf {\bibinfo {volume} {109}},\ \bibinfo {pages} {1492--1505}
  (\bibinfo {year} {1958})}\BibitemShut {NoStop}%
\bibitem [{\citenamefont {Vosk}\ \emph {et~al.}(2015)\citenamefont {Vosk},
  \citenamefont {Huse},\ and\ \citenamefont {Altman}}]{Vosk15}%
  \BibitemOpen
  \bibfield  {author} {\bibinfo {author} {\bibfnamefont {R.}~\bibnamefont
  {Vosk}}, \bibinfo {author} {\bibfnamefont {D.~A.}\ \bibnamefont {Huse}}, \
  and\ \bibinfo {author} {\bibfnamefont {E.}~\bibnamefont {Altman}},\
  }\bibfield  {title} {\enquote {\bibinfo {title} {Theory of the many-body
  localization transition in one-dimensional systems},}\ }\href {\doibase
  10.1103/PhysRevX.5.031032} {\bibfield  {journal} {\bibinfo  {journal} {Phys.
  Rev. X}\ }\textbf {\bibinfo {volume} {5}},\ \bibinfo {pages} {031032}
  (\bibinfo {year} {2015})}\BibitemShut {NoStop}%
\bibitem [{\citenamefont {Altman}\ and\ \citenamefont
  {Vosk}(2015)}]{altman2015review}%
  \BibitemOpen
  \bibfield  {author} {\bibinfo {author} {\bibfnamefont {Ehud}\ \bibnamefont
  {Altman}}\ and\ \bibinfo {author} {\bibfnamefont {Ronen}\ \bibnamefont
  {Vosk}},\ }\bibfield  {title} {\enquote {\bibinfo {title} {Universal dynamics
  and renormalization in many-body-localized systems},}\ }\href {\doibase
  10.1146/annurev-conmatphys-031214-014701} {\bibfield  {journal} {\bibinfo
  {journal} {Annual Review of Condensed Matter Physics}\ }\textbf {\bibinfo
  {volume} {6}},\ \bibinfo {pages} {383--409} (\bibinfo {year} {2015})},\
  \Eprint
  {http://arxiv.org/abs/https://doi.org/10.1146/annurev-conmatphys-031214-014701}
  {https://doi.org/10.1146/annurev-conmatphys-031214-014701} \BibitemShut
  {NoStop}%
\bibitem [{\citenamefont {Agarwal}\ \emph {et~al.}(2015)\citenamefont
  {Agarwal}, \citenamefont {Gopalakrishnan}, \citenamefont {Knap},
  \citenamefont {M\"uller},\ and\ \citenamefont {Demler}}]{Agar15}%
  \BibitemOpen
  \bibfield  {author} {\bibinfo {author} {\bibfnamefont {K.}~\bibnamefont
  {Agarwal}}, \bibinfo {author} {\bibfnamefont {S.}~\bibnamefont
  {Gopalakrishnan}}, \bibinfo {author} {\bibfnamefont {M.}~\bibnamefont
  {Knap}}, \bibinfo {author} {\bibfnamefont {M.}~\bibnamefont {M\"uller}}, \
  and\ \bibinfo {author} {\bibfnamefont {E.}~\bibnamefont {Demler}},\
  }\bibfield  {title} {\enquote {\bibinfo {title} {Anomalous diffusion and
  griffiths effects near the many-body localization transition},}\ }\href
  {\doibase 10.1103/PhysRevLett.114.160401} {\bibfield  {journal} {\bibinfo
  {journal} {Phys. Rev. Lett.}\ }\textbf {\bibinfo {volume} {114}},\ \bibinfo
  {pages} {160401} (\bibinfo {year} {2015})}\BibitemShut {NoStop}%
\bibitem [{\citenamefont {Khemani}\ \emph {et~al.}(2017)\citenamefont
  {Khemani}, \citenamefont {Lim}, \citenamefont {Sheng},\ and\ \citenamefont
  {Huse}}]{Vedi17}%
  \BibitemOpen
  \bibfield  {author} {\bibinfo {author} {\bibfnamefont {V.}~\bibnamefont
  {Khemani}}, \bibinfo {author} {\bibfnamefont {S.~P.}\ \bibnamefont {Lim}},
  \bibinfo {author} {\bibfnamefont {D.~N.}\ \bibnamefont {Sheng}}, \ and\
  \bibinfo {author} {\bibfnamefont {D.~A.}\ \bibnamefont {Huse}},\ }\bibfield
  {title} {\enquote {\bibinfo {title} {Critical properties of the many-body
  localization transition},}\ }\href {\doibase 10.1103/PhysRevX.7.021013}
  {\bibfield  {journal} {\bibinfo  {journal} {Phys. Rev. X}\ }\textbf {\bibinfo
  {volume} {7}},\ \bibinfo {pages} {021013} (\bibinfo {year}
  {2017})}\BibitemShut {NoStop}%
\bibitem [{\citenamefont {Pietracaprina}\ \emph {et~al.}(2017)\citenamefont
  {Pietracaprina}, \citenamefont {Parisi}, \citenamefont {Mariano},
  \citenamefont {Pascazio},\ and\ \citenamefont
  {Scardicchio}}]{pietracaprina2017entanglement}%
  \BibitemOpen
  \bibfield  {author} {\bibinfo {author} {\bibfnamefont {F.}~\bibnamefont
  {Pietracaprina}}, \bibinfo {author} {\bibfnamefont {G.}~\bibnamefont
  {Parisi}}, \bibinfo {author} {\bibfnamefont {A.}~\bibnamefont {Mariano}},
  \bibinfo {author} {\bibfnamefont {S.}~\bibnamefont {Pascazio}}, \ and\
  \bibinfo {author} {\bibfnamefont {A.}~\bibnamefont {Scardicchio}},\
  }\bibfield  {title} {\enquote {\bibinfo {title} {Entanglement critical length
  at the many-body localization transition},}\ }\href
  {http://iopscience.iop.org/article/10.1088/1742-5468/aa9338/meta} {\bibfield
  {journal} {\bibinfo  {journal} {J. Stat. Mech.: Theory and Experiment}\
  }\textbf {\bibinfo {volume} {2017}},\ \bibinfo {pages} {113102} (\bibinfo
  {year} {2017})}\BibitemShut {NoStop}%
\bibitem [{\citenamefont {Doggen}\ \emph {et~al.}(2018)\citenamefont {Doggen},
  \citenamefont {Schindler}, \citenamefont {Tikhonov}, \citenamefont {Mirlin},
  \citenamefont {Neupert}, \citenamefont {Polyakov},\ and\ \citenamefont
  {Gornyi}}]{Tik_MPS}%
  \BibitemOpen
  \bibfield  {author} {\bibinfo {author} {\bibfnamefont {E.~V.~H.}\
  \bibnamefont {Doggen}}, \bibinfo {author} {\bibfnamefont {F.}~\bibnamefont
  {Schindler}}, \bibinfo {author} {\bibfnamefont {K.~S.}\ \bibnamefont
  {Tikhonov}}, \bibinfo {author} {\bibfnamefont {A.~D.}\ \bibnamefont
  {Mirlin}}, \bibinfo {author} {\bibfnamefont {T.}~\bibnamefont {Neupert}},
  \bibinfo {author} {\bibfnamefont {D.~G.}\ \bibnamefont {Polyakov}}, \ and\
  \bibinfo {author} {\bibfnamefont {I.~V.}\ \bibnamefont {Gornyi}},\ }\bibfield
   {title} {\enquote {\bibinfo {title} {Many-body localization and
  delocalization in large quantum chains},}\ }\href {\doibase
  10.1103/PhysRevB.98.174202} {\bibfield  {journal} {\bibinfo  {journal} {Phys.
  Rev. B}\ }\textbf {\bibinfo {volume} {98}},\ \bibinfo {pages} {174202}
  (\bibinfo {year} {2018})}\BibitemShut {NoStop}%
\bibitem [{\citenamefont {{\v{S}}untajs}\ \emph {et~al.}(2019)\citenamefont
  {{\v{S}}untajs}, \citenamefont {Bon{\v{c}}a}, \citenamefont {Prosen},\ and\
  \citenamefont {Vidmar}}]{vsuntajs2019quantum}%
  \BibitemOpen
  \bibfield  {author} {\bibinfo {author} {\bibfnamefont {J.}~\bibnamefont
  {{\v{S}}untajs}}, \bibinfo {author} {\bibfnamefont {J.}~\bibnamefont
  {Bon{\v{c}}a}}, \bibinfo {author} {\bibfnamefont {T.}~\bibnamefont {Prosen}},
  \ and\ \bibinfo {author} {\bibfnamefont {L.}~\bibnamefont {Vidmar}},\
  }\href@noop {} {\enquote {\bibinfo {title} {Quantum chaos challenges
  many-body localization},}\ } (\bibinfo {year} {2019}),\ \Eprint
  {http://arxiv.org/abs/1905.06345} {arXiv:1905.06345} \BibitemShut {NoStop}%
\bibitem [{\citenamefont {Chandran}\ \emph
  {et~al.}(2015{\natexlab{a}})\citenamefont {Chandran}, \citenamefont
  {Laumann},\ and\ \citenamefont {Oganesyan}}]{chandran2015finite}%
  \BibitemOpen
  \bibfield  {author} {\bibinfo {author} {\bibfnamefont {A}~\bibnamefont
  {Chandran}}, \bibinfo {author} {\bibfnamefont {CR}~\bibnamefont {Laumann}}, \
  and\ \bibinfo {author} {\bibfnamefont {V}~\bibnamefont {Oganesyan}},\
  }\href@noop {} {\enquote {\bibinfo {title} {Finite size scaling bounds on
  many-body localized phase transitions},}\ } (\bibinfo {year}
  {2015}{\natexlab{a}}),\ \Eprint {http://arxiv.org/abs/1509.04285}
  {arXiv:1509.04285} \BibitemShut {NoStop}%
\bibitem [{\citenamefont {Luitz}\ and\ \citenamefont
  {Bar~Lev}(2016)}]{Luitz16}%
  \BibitemOpen
  \bibfield  {author} {\bibinfo {author} {\bibfnamefont {D.~J.}\ \bibnamefont
  {Luitz}}\ and\ \bibinfo {author} {\bibfnamefont {Y.}~\bibnamefont
  {Bar~Lev}},\ }\bibfield  {title} {\enquote {\bibinfo {title} {Anomalous
  thermalization in ergodic systems},}\ }\href {\doibase
  10.1103/PhysRevLett.117.170404} {\bibfield  {journal} {\bibinfo  {journal}
  {Phys. Rev. Lett.}\ }\textbf {\bibinfo {volume} {117}},\ \bibinfo {pages}
  {170404} (\bibinfo {year} {2016})}\BibitemShut {NoStop}%
\bibitem [{\citenamefont {Bar~Lev}\ \emph {et~al.}(2015)\citenamefont
  {Bar~Lev}, \citenamefont {Cohen},\ and\ \citenamefont {Reichman}}]{Bar15}%
  \BibitemOpen
  \bibfield  {author} {\bibinfo {author} {\bibfnamefont {Y.}~\bibnamefont
  {Bar~Lev}}, \bibinfo {author} {\bibfnamefont {G.}~\bibnamefont {Cohen}}, \
  and\ \bibinfo {author} {\bibfnamefont {D.~R.}\ \bibnamefont {Reichman}},\
  }\bibfield  {title} {\enquote {\bibinfo {title} {Absence of diffusion in an
  interacting system of spinless fermions on a one-dimensional disordered
  lattice},}\ }\href {\doibase 10.1103/PhysRevLett.114.100601} {\bibfield
  {journal} {\bibinfo  {journal} {Phys. Rev. Lett.}\ }\textbf {\bibinfo
  {volume} {114}},\ \bibinfo {pages} {100601} (\bibinfo {year}
  {2015})}\BibitemShut {NoStop}%
\bibitem [{\citenamefont {Prelov\ifmmode~\check{s}\else \v{s}\fi{}ek}\ and\
  \citenamefont {Herbrych}(2017)}]{Pre17}%
  \BibitemOpen
  \bibfield  {author} {\bibinfo {author} {\bibfnamefont {P.}~\bibnamefont
  {Prelov\ifmmode~\check{s}\else \v{s}\fi{}ek}}\ and\ \bibinfo {author}
  {\bibfnamefont {J.}~\bibnamefont {Herbrych}},\ }\bibfield  {title} {\enquote
  {\bibinfo {title} {Self-consistent approach to many-body localization and
  subdiffusion},}\ }\href {\doibase 10.1103/PhysRevB.96.035130} {\bibfield
  {journal} {\bibinfo  {journal} {Phys. Rev. B}\ }\textbf {\bibinfo {volume}
  {96}},\ \bibinfo {pages} {035130} (\bibinfo {year} {2017})}\BibitemShut
  {NoStop}%
\bibitem [{\citenamefont {{Kerala Varma}}\ \emph {et~al.}(2015)\citenamefont
  {{Kerala Varma}}, \citenamefont {{Lerose}}, \citenamefont {{Pietracaprina}},
  \citenamefont {{Goold}},\ and\ \citenamefont {{Scardicchio}}}]{Vipin2015}%
  \BibitemOpen
  \bibfield  {author} {\bibinfo {author} {\bibfnamefont {V.}~\bibnamefont
  {{Kerala Varma}}}, \bibinfo {author} {\bibfnamefont {A.}~\bibnamefont
  {{Lerose}}}, \bibinfo {author} {\bibfnamefont {F.}~\bibnamefont
  {{Pietracaprina}}}, \bibinfo {author} {\bibfnamefont {J.}~\bibnamefont
  {{Goold}}}, \ and\ \bibinfo {author} {\bibfnamefont {A.}~\bibnamefont
  {{Scardicchio}}},\ }\href@noop {} {\enquote {\bibinfo {title} {Energy
  diffusion in the ergodic phase of a many body localizable spin chain},}\ }
  (\bibinfo {year} {2015}),\ \Eprint {http://arxiv.org/abs/1511.09144}
  {arXiv:1511.09144} \BibitemShut {NoStop}%
\bibitem [{\citenamefont {\ifmmode \check{Z}\else
  \v{Z}\fi{}nidari\ifmmode~\check{c}\else \v{c}\fi{}}\ \emph
  {et~al.}(2016)\citenamefont {\ifmmode \check{Z}\else
  \v{Z}\fi{}nidari\ifmmode~\check{c}\else \v{c}\fi{}}, \citenamefont
  {Scardicchio},\ and\ \citenamefont {Varma}}]{Scardi16}%
  \BibitemOpen
  \bibfield  {author} {\bibinfo {author} {\bibfnamefont {M.}~\bibnamefont
  {\ifmmode \check{Z}\else \v{Z}\fi{}nidari\ifmmode~\check{c}\else
  \v{c}\fi{}}}, \bibinfo {author} {\bibfnamefont {A.}~\bibnamefont
  {Scardicchio}}, \ and\ \bibinfo {author} {\bibfnamefont {V.~K.}\ \bibnamefont
  {Varma}},\ }\bibfield  {title} {\enquote {\bibinfo {title} {Diffusive and
  subdiffusive spin transport in the ergodic phase of a many-body localizable
  system},}\ }\href {\doibase 10.1103/PhysRevLett.117.040601} {\bibfield
  {journal} {\bibinfo  {journal} {Phys. Rev. Lett.}\ }\textbf {\bibinfo
  {volume} {117}},\ \bibinfo {pages} {040601} (\bibinfo {year}
  {2016})}\BibitemShut {NoStop}%
\bibitem [{\citenamefont {Gopalakrishnan}\ \emph {et~al.}(2016)\citenamefont
  {Gopalakrishnan}, \citenamefont {Agarwal}, \citenamefont {Demler},
  \citenamefont {Huse},\ and\ \citenamefont {Knap}}]{Gopa16}%
  \BibitemOpen
  \bibfield  {author} {\bibinfo {author} {\bibfnamefont {S.}~\bibnamefont
  {Gopalakrishnan}}, \bibinfo {author} {\bibfnamefont {K.}~\bibnamefont
  {Agarwal}}, \bibinfo {author} {\bibfnamefont {E.~A.}\ \bibnamefont {Demler}},
  \bibinfo {author} {\bibfnamefont {D.~A.}\ \bibnamefont {Huse}}, \ and\
  \bibinfo {author} {\bibfnamefont {M.}~\bibnamefont {Knap}},\ }\bibfield
  {title} {\enquote {\bibinfo {title} {Griffiths effects and slow dynamics in
  nearly many-body localized systems},}\ }\href {\doibase
  10.1103/PhysRevB.93.134206} {\bibfield  {journal} {\bibinfo  {journal} {Phys.
  Rev. B}\ }\textbf {\bibinfo {volume} {93}},\ \bibinfo {pages} {134206}
  (\bibinfo {year} {2016})}\BibitemShut {NoStop}%
\bibitem [{\citenamefont {Khait}\ \emph {et~al.}(2016)\citenamefont {Khait},
  \citenamefont {Gazit}, \citenamefont {Yao},\ and\ \citenamefont
  {Auerbach}}]{Khait16}%
  \BibitemOpen
  \bibfield  {author} {\bibinfo {author} {\bibfnamefont {Ilia}\ \bibnamefont
  {Khait}}, \bibinfo {author} {\bibfnamefont {Snir}\ \bibnamefont {Gazit}},
  \bibinfo {author} {\bibfnamefont {Norman~Y.}\ \bibnamefont {Yao}}, \ and\
  \bibinfo {author} {\bibfnamefont {Assa}\ \bibnamefont {Auerbach}},\
  }\bibfield  {title} {\enquote {\bibinfo {title} {Spin transport of weakly
  disordered heisenberg chain at infinite temperature},}\ }\href {\doibase
  10.1103/PhysRevB.93.224205} {\bibfield  {journal} {\bibinfo  {journal} {Phys.
  Rev. B}\ }\textbf {\bibinfo {volume} {93}},\ \bibinfo {pages} {224205}
  (\bibinfo {year} {2016})}\BibitemShut {NoStop}%
\bibitem [{\citenamefont {Schulz}\ \emph {et~al.}(2019)\citenamefont {Schulz},
  \citenamefont {Taylor}, \citenamefont {Scardicchio},\ and\ \citenamefont
  {Znidaric}}]{schulz2019phenomenology}%
  \BibitemOpen
  \bibfield  {author} {\bibinfo {author} {\bibfnamefont {Maximilian}\
  \bibnamefont {Schulz}}, \bibinfo {author} {\bibfnamefont {Scott}\
  \bibnamefont {Taylor}}, \bibinfo {author} {\bibfnamefont {Antonello}\
  \bibnamefont {Scardicchio}}, \ and\ \bibinfo {author} {\bibfnamefont {Marko}\
  \bibnamefont {Znidaric}},\ }\href@noop {} {\enquote {\bibinfo {title}
  {Phenomenology of anomalous transport in disordered one-dimensional
  systems},}\ } (\bibinfo {year} {2019}),\ \Eprint
  {http://arxiv.org/abs/1909.09507} {arXiv:1909.09507} \BibitemShut {NoStop}%
\bibitem [{\citenamefont {Bera}\ \emph {et~al.}(2017)\citenamefont {Bera},
  \citenamefont {De~Tomasi}, \citenamefont {Weiner},\ and\ \citenamefont
  {Evers}}]{Bera17}%
  \BibitemOpen
  \bibfield  {author} {\bibinfo {author} {\bibfnamefont {S.}~\bibnamefont
  {Bera}}, \bibinfo {author} {\bibfnamefont {G.}~\bibnamefont {De~Tomasi}},
  \bibinfo {author} {\bibfnamefont {F.}~\bibnamefont {Weiner}}, \ and\ \bibinfo
  {author} {\bibfnamefont {F.}~\bibnamefont {Evers}},\ }\bibfield  {title}
  {\enquote {\bibinfo {title} {Density propagator for many-body localization:
  Finite-size effects, transient subdiffusion, and exponential decay},}\ }\href
  {\doibase 10.1103/PhysRevLett.118.196801} {\bibfield  {journal} {\bibinfo
  {journal} {Phys. Rev. Lett.}\ }\textbf {\bibinfo {volume} {118}},\ \bibinfo
  {pages} {196801} (\bibinfo {year} {2017})}\BibitemShut {NoStop}%
\bibitem [{\citenamefont {Gopalakrishnan}\ \emph {et~al.}(2015)\citenamefont
  {Gopalakrishnan}, \citenamefont {M\"uller}, \citenamefont {Khemani},
  \citenamefont {Knap}, \citenamefont {Demler},\ and\ \citenamefont
  {Huse}}]{gopalakrishnan2015low}%
  \BibitemOpen
  \bibfield  {author} {\bibinfo {author} {\bibfnamefont {Sarang}\ \bibnamefont
  {Gopalakrishnan}}, \bibinfo {author} {\bibfnamefont {Markus}\ \bibnamefont
  {M\"uller}}, \bibinfo {author} {\bibfnamefont {Vedika}\ \bibnamefont
  {Khemani}}, \bibinfo {author} {\bibfnamefont {Michael}\ \bibnamefont {Knap}},
  \bibinfo {author} {\bibfnamefont {Eugene}\ \bibnamefont {Demler}}, \ and\
  \bibinfo {author} {\bibfnamefont {David~A.}\ \bibnamefont {Huse}},\
  }\bibfield  {title} {\enquote {\bibinfo {title} {Low-frequency conductivity
  in many-body localized systems},}\ }\href {\doibase
  10.1103/PhysRevB.92.104202} {\bibfield  {journal} {\bibinfo  {journal} {Phys.
  Rev. B}\ }\textbf {\bibinfo {volume} {92}},\ \bibinfo {pages} {104202}
  (\bibinfo {year} {2015})}\BibitemShut {NoStop}%
\bibitem [{\citenamefont {Weiner}\ \emph {et~al.}(2019)\citenamefont {Weiner},
  \citenamefont {Evers},\ and\ \citenamefont {Bera}}]{Weiner19}%
  \BibitemOpen
  \bibfield  {author} {\bibinfo {author} {\bibfnamefont {Felix}\ \bibnamefont
  {Weiner}}, \bibinfo {author} {\bibfnamefont {Ferdinand}\ \bibnamefont
  {Evers}}, \ and\ \bibinfo {author} {\bibfnamefont {Soumya}\ \bibnamefont
  {Bera}},\ }\bibfield  {title} {\enquote {\bibinfo {title} {Slow dynamics and
  strong finite-size effects in many-body localization with random and
  quasiperiodic potentials},}\ }\href {\doibase 10.1103/PhysRevB.100.104204}
  {\bibfield  {journal} {\bibinfo  {journal} {Phys. Rev. B}\ }\textbf {\bibinfo
  {volume} {100}},\ \bibinfo {pages} {104204} (\bibinfo {year}
  {2019})}\BibitemShut {NoStop}%
\bibitem [{\citenamefont {De~Tomasi}\ \emph
  {et~al.}(2019{\natexlab{a}})\citenamefont {De~Tomasi}, \citenamefont
  {Hetterich}, \citenamefont {Sala},\ and\ \citenamefont
  {Pollmann}}]{detomasi2019dynamics}%
  \BibitemOpen
  \bibfield  {author} {\bibinfo {author} {\bibfnamefont {Giuseppe}\
  \bibnamefont {De~Tomasi}}, \bibinfo {author} {\bibfnamefont {Daniel}\
  \bibnamefont {Hetterich}}, \bibinfo {author} {\bibfnamefont {Pablo}\
  \bibnamefont {Sala}}, \ and\ \bibinfo {author} {\bibfnamefont {Frank}\
  \bibnamefont {Pollmann}},\ }\href@noop {} {\enquote {\bibinfo {title}
  {Dynamics of strongly interacting systems: From fock-space fragmentation to
  many-body localization},}\ } (\bibinfo {year} {2019}{\natexlab{a}}),\ \Eprint
  {http://arxiv.org/abs/1909.03073} {arXiv:1909.03073} \BibitemShut {NoStop}%
\bibitem [{\citenamefont {Khemani}\ \emph {et~al.}(2016)\citenamefont
  {Khemani}, \citenamefont {Pollmann},\ and\ \citenamefont
  {Sondhi}}]{Khemani2015obtaining}%
  \BibitemOpen
  \bibfield  {author} {\bibinfo {author} {\bibfnamefont {Vedika}\ \bibnamefont
  {Khemani}}, \bibinfo {author} {\bibfnamefont {Frank}\ \bibnamefont
  {Pollmann}}, \ and\ \bibinfo {author} {\bibfnamefont {S.~L.}\ \bibnamefont
  {Sondhi}},\ }\bibfield  {title} {\enquote {\bibinfo {title} {Obtaining highly
  excited eigenstates of many-body localized hamiltonians by the density matrix
  renormalization group approach},}\ }\href {\doibase
  10.1103/PhysRevLett.116.247204} {\bibfield  {journal} {\bibinfo  {journal}
  {Phys. Rev. Lett.}\ }\textbf {\bibinfo {volume} {116}},\ \bibinfo {pages}
  {247204} (\bibinfo {year} {2016})}\BibitemShut {NoStop}%
\bibitem [{\citenamefont {Chandran}\ \emph
  {et~al.}(2015{\natexlab{b}})\citenamefont {Chandran}, \citenamefont
  {Carrasquilla}, \citenamefont {Kim}, \citenamefont {Abanin},\ and\
  \citenamefont {Vidal}}]{Chandran2015spectral}%
  \BibitemOpen
  \bibfield  {author} {\bibinfo {author} {\bibfnamefont {A.}~\bibnamefont
  {Chandran}}, \bibinfo {author} {\bibfnamefont {J.}~\bibnamefont
  {Carrasquilla}}, \bibinfo {author} {\bibfnamefont {I.~H.}\ \bibnamefont
  {Kim}}, \bibinfo {author} {\bibfnamefont {D.~A.}\ \bibnamefont {Abanin}}, \
  and\ \bibinfo {author} {\bibfnamefont {G.}~\bibnamefont {Vidal}},\ }\bibfield
   {title} {\enquote {\bibinfo {title} {Spectral tensor networks for many-body
  localization},}\ }\href {\doibase 10.1103/PhysRevB.92.024201} {\bibfield
  {journal} {\bibinfo  {journal} {Phys. Rev. B}\ }\textbf {\bibinfo {volume}
  {92}},\ \bibinfo {pages} {024201} (\bibinfo {year}
  {2015}{\natexlab{b}})}\BibitemShut {NoStop}%
\bibitem [{\citenamefont {Yu}\ \emph {et~al.}(2017)\citenamefont {Yu},
  \citenamefont {Pekker},\ and\ \citenamefont {Clark}}]{YuMPS2015}%
  \BibitemOpen
  \bibfield  {author} {\bibinfo {author} {\bibfnamefont {Xiongjie}\
  \bibnamefont {Yu}}, \bibinfo {author} {\bibfnamefont {David}\ \bibnamefont
  {Pekker}}, \ and\ \bibinfo {author} {\bibfnamefont {Bryan~K.}\ \bibnamefont
  {Clark}},\ }\bibfield  {title} {\enquote {\bibinfo {title} {Finding matrix
  product state representations of highly excited eigenstates of many-body
  localized hamiltonians},}\ }\href {\doibase 10.1103/PhysRevLett.118.017201}
  {\bibfield  {journal} {\bibinfo  {journal} {Phys. Rev. Lett.}\ }\textbf
  {\bibinfo {volume} {118}},\ \bibinfo {pages} {017201} (\bibinfo {year}
  {2017})}\BibitemShut {NoStop}%
\bibitem [{\citenamefont {Lim}\ and\ \citenamefont {Sheng}(2016)}]{LimMPS2015}%
  \BibitemOpen
  \bibfield  {author} {\bibinfo {author} {\bibfnamefont {S.~P.}\ \bibnamefont
  {Lim}}\ and\ \bibinfo {author} {\bibfnamefont {D.~N.}\ \bibnamefont
  {Sheng}},\ }\bibfield  {title} {\enquote {\bibinfo {title} {Many-body
  localization and transition by density matrix renormalization group and exact
  diagonalization studies},}\ }\href {\doibase 10.1103/PhysRevB.94.045111}
  {\bibfield  {journal} {\bibinfo  {journal} {Phys. Rev. B}\ }\textbf {\bibinfo
  {volume} {94}},\ \bibinfo {pages} {045111} (\bibinfo {year}
  {2016})}\BibitemShut {NoStop}%
\bibitem [{\citenamefont {Biroli}\ and\ \citenamefont
  {Tarzia}(2017)}]{Biroli2017Dynamics}%
  \BibitemOpen
  \bibfield  {author} {\bibinfo {author} {\bibfnamefont {G.}~\bibnamefont
  {Biroli}}\ and\ \bibinfo {author} {\bibfnamefont {M.}~\bibnamefont
  {Tarzia}},\ }\bibfield  {title} {\enquote {\bibinfo {title} {Delocalized
  glassy dynamics and many-body localization},}\ }\href {\doibase
  10.1103/PhysRevB.96.201114} {\bibfield  {journal} {\bibinfo  {journal} {Phys.
  Rev. B}\ }\textbf {\bibinfo {volume} {96}},\ \bibinfo {pages} {201114(R)}
  (\bibinfo {year} {2017})}\BibitemShut {NoStop}%
\bibitem [{\citenamefont {Tikhonov}\ and\ \citenamefont
  {Mirlin}(2019{\natexlab{a}})}]{Tikh2019Critical}%
  \BibitemOpen
  \bibfield  {author} {\bibinfo {author} {\bibfnamefont {K.~S.}\ \bibnamefont
  {Tikhonov}}\ and\ \bibinfo {author} {\bibfnamefont {A.~D.}\ \bibnamefont
  {Mirlin}},\ }\bibfield  {title} {\enquote {\bibinfo {title} {Critical
  behavior at the localization transition on random regular graphs},}\ }\href
  {\doibase 10.1103/PhysRevB.99.214202} {\bibfield  {journal} {\bibinfo
  {journal} {Phys. Rev. B}\ }\textbf {\bibinfo {volume} {99}},\ \bibinfo
  {pages} {214202} (\bibinfo {year} {2019}{\natexlab{a}})}\BibitemShut
  {NoStop}%
\bibitem [{\citenamefont {{Avetisov}}\ \emph {et~al.}(2016)\citenamefont
  {{Avetisov}}, \citenamefont {{Gorsky}}, \citenamefont {{Nechaev}},\ and\
  \citenamefont {{Valba}}}]{Avetisov16}%
  \BibitemOpen
  \bibfield  {author} {\bibinfo {author} {\bibfnamefont {V.}~\bibnamefont
  {{Avetisov}}}, \bibinfo {author} {\bibfnamefont {A.}~\bibnamefont
  {{Gorsky}}}, \bibinfo {author} {\bibfnamefont {S.}~\bibnamefont {{Nechaev}}},
  \ and\ \bibinfo {author} {\bibfnamefont {O.}~\bibnamefont {{Valba}}},\
  }\href@noop {} {\enquote {\bibinfo {title} {{Many-body localization and new
  critical phenomena in regular random graphs and constrained Erd{\H
  o}s-R{\'e}nyi networks}},}\ } (\bibinfo {year} {2016}),\ \Eprint
  {http://arxiv.org/abs/1611.08531} {arXiv:1611.08531} \BibitemShut {NoStop}%
\bibitem [{\citenamefont {Logan}\ and\ \citenamefont {Welsh}(2019)}]{Logan19}%
  \BibitemOpen
  \bibfield  {author} {\bibinfo {author} {\bibfnamefont {D.~E.}\ \bibnamefont
  {Logan}}\ and\ \bibinfo {author} {\bibfnamefont {S.}~\bibnamefont {Welsh}},\
  }\bibfield  {title} {\enquote {\bibinfo {title} {Many-body localization in
  fock space: A local perspective},}\ }\href {\doibase
  10.1103/PhysRevB.99.045131} {\bibfield  {journal} {\bibinfo  {journal} {Phys.
  Rev. B}\ }\textbf {\bibinfo {volume} {99}},\ \bibinfo {pages} {045131}
  (\bibinfo {year} {2019})}\BibitemShut {NoStop}%
\bibitem [{\citenamefont {Roy}\ \emph {et~al.}(2019{\natexlab{a}})\citenamefont
  {Roy}, \citenamefont {Logan},\ and\ \citenamefont {Chalker}}]{Roy1}%
  \BibitemOpen
  \bibfield  {author} {\bibinfo {author} {\bibfnamefont {S.}~\bibnamefont
  {Roy}}, \bibinfo {author} {\bibfnamefont {D.~E.}\ \bibnamefont {Logan}}, \
  and\ \bibinfo {author} {\bibfnamefont {J.~T.}\ \bibnamefont {Chalker}},\
  }\bibfield  {title} {\enquote {\bibinfo {title} {Exact solution of a
  percolation analog for the many-body localization transition},}\ }\href
  {\doibase 10.1103/PhysRevB.99.220201} {\bibfield  {journal} {\bibinfo
  {journal} {Phys. Rev. B}\ }\textbf {\bibinfo {volume} {99}},\ \bibinfo
  {pages} {220201(R)} (\bibinfo {year} {2019}{\natexlab{a}})}\BibitemShut
  {NoStop}%
\bibitem [{\citenamefont {Roy}\ \emph {et~al.}(2019{\natexlab{b}})\citenamefont
  {Roy}, \citenamefont {Chalker},\ and\ \citenamefont {Logan}}]{Roy2}%
  \BibitemOpen
  \bibfield  {author} {\bibinfo {author} {\bibfnamefont {S.}~\bibnamefont
  {Roy}}, \bibinfo {author} {\bibfnamefont {J.~T.}\ \bibnamefont {Chalker}}, \
  and\ \bibinfo {author} {\bibfnamefont {D.~E.}\ \bibnamefont {Logan}},\
  }\bibfield  {title} {\enquote {\bibinfo {title} {Percolation in fock space as
  a proxy for many-body localization},}\ }\href {\doibase
  10.1103/PhysRevB.99.104206} {\bibfield  {journal} {\bibinfo  {journal} {Phys.
  Rev. B}\ }\textbf {\bibinfo {volume} {99}},\ \bibinfo {pages} {104206}
  (\bibinfo {year} {2019}{\natexlab{b}})}\BibitemShut {NoStop}%
\bibitem [{\citenamefont {Altshuler}\ \emph {et~al.}(1997)\citenamefont
  {Altshuler}, \citenamefont {Gefen}, \citenamefont {Kamenev},\ and\
  \citenamefont {Levitov}}]{Alt97}%
  \BibitemOpen
  \bibfield  {author} {\bibinfo {author} {\bibfnamefont {B.~L.}\ \bibnamefont
  {Altshuler}}, \bibinfo {author} {\bibfnamefont {Y.}~\bibnamefont {Gefen}},
  \bibinfo {author} {\bibfnamefont {A.}~\bibnamefont {Kamenev}}, \ and\
  \bibinfo {author} {\bibfnamefont {L.~S.}\ \bibnamefont {Levitov}},\
  }\bibfield  {title} {\enquote {\bibinfo {title} {Quasiparticle lifetime in a
  finite system: A nonperturbative approach},}\ }\href {\doibase
  10.1103/PhysRevLett.78.2803} {\bibfield  {journal} {\bibinfo  {journal}
  {Phys. Rev. Lett.}\ }\textbf {\bibinfo {volume} {78}},\ \bibinfo {pages}
  {2803--2806} (\bibinfo {year} {1997})}\BibitemShut {NoStop}%
\bibitem [{\citenamefont {Abou-Chacra}\ \emph {et~al.}(1973)\citenamefont
  {Abou-Chacra}, \citenamefont {Thouless},\ and\ \citenamefont
  {Anderson}}]{Abou73}%
  \BibitemOpen
  \bibfield  {author} {\bibinfo {author} {\bibfnamefont {R}~\bibnamefont
  {Abou-Chacra}}, \bibinfo {author} {\bibfnamefont {D~J}\ \bibnamefont
  {Thouless}}, \ and\ \bibinfo {author} {\bibfnamefont {P~W}\ \bibnamefont
  {Anderson}},\ }\bibfield  {title} {\enquote {\bibinfo {title} {A
  selfconsistent theory of localization},}\ }\href
  {http://stacks.iop.org/0022-3719/6/i=10/a=009} {\bibfield  {journal}
  {\bibinfo  {journal} {J. Phys. C: Solid State Physics}\ }\textbf {\bibinfo
  {volume} {6}},\ \bibinfo {pages} {1734} (\bibinfo {year} {1973})}\BibitemShut
  {NoStop}%
\bibitem [{\citenamefont {De~Luca}\ \emph {et~al.}(2014)\citenamefont
  {De~Luca}, \citenamefont {Altshuler}, \citenamefont {Kravtsov},\ and\
  \citenamefont {Scardicchio}}]{Deluca14}%
  \BibitemOpen
  \bibfield  {author} {\bibinfo {author} {\bibfnamefont {A.}~\bibnamefont
  {De~Luca}}, \bibinfo {author} {\bibfnamefont {B.~L.}\ \bibnamefont
  {Altshuler}}, \bibinfo {author} {\bibfnamefont {V.~E.}\ \bibnamefont
  {Kravtsov}}, \ and\ \bibinfo {author} {\bibfnamefont {A.}~\bibnamefont
  {Scardicchio}},\ }\bibfield  {title} {\enquote {\bibinfo {title} {Anderson
  localization on the bethe lattice: Nonergodicity of extended states},}\
  }\href {\doibase 10.1103/PhysRevLett.113.046806} {\bibfield  {journal}
  {\bibinfo  {journal} {Phys. Rev. Lett.}\ }\textbf {\bibinfo {volume} {113}},\
  \bibinfo {pages} {046806} (\bibinfo {year} {2014})}\BibitemShut {NoStop}%
\bibitem [{\citenamefont {Altshuler}\ \emph {et~al.}(2016)\citenamefont
  {Altshuler}, \citenamefont {Cuevas}, \citenamefont {Ioffe},\ and\
  \citenamefont {Kravtsov}}]{Alt16}%
  \BibitemOpen
  \bibfield  {author} {\bibinfo {author} {\bibfnamefont {B.~L.}\ \bibnamefont
  {Altshuler}}, \bibinfo {author} {\bibfnamefont {E.}~\bibnamefont {Cuevas}},
  \bibinfo {author} {\bibfnamefont {L.~B.}\ \bibnamefont {Ioffe}}, \ and\
  \bibinfo {author} {\bibfnamefont {V.~E.}\ \bibnamefont {Kravtsov}},\
  }\bibfield  {title} {\enquote {\bibinfo {title} {Nonergodic phases in
  strongly disordered random regular graphs},}\ }\href {\doibase
  10.1103/PhysRevLett.117.156601} {\bibfield  {journal} {\bibinfo  {journal}
  {Phys. Rev. Lett.}\ }\textbf {\bibinfo {volume} {117}},\ \bibinfo {pages}
  {156601} (\bibinfo {year} {2016})}\BibitemShut {NoStop}%
\bibitem [{\citenamefont {{Altshuler}}\ \emph {et~al.}(2016)\citenamefont
  {{Altshuler}}, \citenamefont {{Ioffe}},\ and\ \citenamefont
  {{Kravtsov}}}]{Alts16}%
  \BibitemOpen
  \bibfield  {author} {\bibinfo {author} {\bibfnamefont {B.~L.}\ \bibnamefont
  {{Altshuler}}}, \bibinfo {author} {\bibfnamefont {L.~B.}\ \bibnamefont
  {{Ioffe}}}, \ and\ \bibinfo {author} {\bibfnamefont {V.~E.}\ \bibnamefont
  {{Kravtsov}}},\ }\href@noop {} {\enquote {\bibinfo {title} {{Multifractal
  states in self-consistent theory of localization: analytical solution}},}\ }
  (\bibinfo {year} {2016}),\ \Eprint {http://arxiv.org/abs/1610.00758}
  {arXiv:1610.00758} \BibitemShut {NoStop}%
\bibitem [{\citenamefont {Kravtsov}\ \emph {et~al.}(2018)\citenamefont
  {Kravtsov}, \citenamefont {Altshuler},\ and\ \citenamefont {Ioffe}}]{Kra18}%
  \BibitemOpen
  \bibfield  {author} {\bibinfo {author} {\bibfnamefont {V.E.}\ \bibnamefont
  {Kravtsov}}, \bibinfo {author} {\bibfnamefont {B.L.}\ \bibnamefont
  {Altshuler}}, \ and\ \bibinfo {author} {\bibfnamefont {L.B.}\ \bibnamefont
  {Ioffe}},\ }\bibfield  {title} {\enquote {\bibinfo {title} {Non-ergodic
  delocalized phase in anderson model on bethe lattice and regular graph},}\
  }\href {\doibase https://doi.org/10.1016/j.aop.2017.12.009} {\bibfield
  {journal} {\bibinfo  {journal} {Annals of Physics}\ }\textbf {\bibinfo
  {volume} {389}},\ \bibinfo {pages} {148 -- 191} (\bibinfo {year}
  {2018})}\BibitemShut {NoStop}%
\bibitem [{\citenamefont {Biroli}\ \emph {et~al.}(2010)\citenamefont {Biroli},
  \citenamefont {Semerjian},\ and\ \citenamefont {Tarzia}}]{Biroli:2010wi}%
  \BibitemOpen
  \bibfield  {author} {\bibinfo {author} {\bibfnamefont {G.}~\bibnamefont
  {Biroli}}, \bibinfo {author} {\bibfnamefont {G.}~\bibnamefont {Semerjian}}, \
  and\ \bibinfo {author} {\bibfnamefont {M.}~\bibnamefont {Tarzia}},\
  }\bibfield  {title} {\enquote {\bibinfo {title} {{Anderson model on {Bethe}
  lattices: density of states, localization properties and isolated
  eigenvalue}},}\ }\href {https://doi.org/10.1143/PTPS.184.187} {\bibfield
  {journal} {\bibinfo  {journal} {Prog. Theor. Phys. Suppl.}\ }\textbf
  {\bibinfo {volume} {184}},\ \bibinfo {pages} {187} (\bibinfo {year}
  {2010})}\BibitemShut {NoStop}%
\bibitem [{\citenamefont {Garc\'{\i}a-Mata}\ \emph {et~al.}(2017)\citenamefont
  {Garc\'{\i}a-Mata}, \citenamefont {Giraud}, \citenamefont {Georgeot},
  \citenamefont {Martin}, \citenamefont {Dubertrand},\ and\ \citenamefont
  {Lemari\'e}}]{Lemarie17Small_K}%
  \BibitemOpen
  \bibfield  {author} {\bibinfo {author} {\bibfnamefont {I.}~\bibnamefont
  {Garc\'{\i}a-Mata}}, \bibinfo {author} {\bibfnamefont {O.}~\bibnamefont
  {Giraud}}, \bibinfo {author} {\bibfnamefont {B.}~\bibnamefont {Georgeot}},
  \bibinfo {author} {\bibfnamefont {J.}~\bibnamefont {Martin}}, \bibinfo
  {author} {\bibfnamefont {R.}~\bibnamefont {Dubertrand}}, \ and\ \bibinfo
  {author} {\bibfnamefont {G.}~\bibnamefont {Lemari\'e}},\ }\bibfield  {title}
  {\enquote {\bibinfo {title} {Scaling theory of the anderson transition in
  random graphs: Ergodicity and universality},}\ }\href {\doibase
  10.1103/PhysRevLett.118.166801} {\bibfield  {journal} {\bibinfo  {journal}
  {Phys. Rev. Lett.}\ }\textbf {\bibinfo {volume} {118}},\ \bibinfo {pages}
  {166801} (\bibinfo {year} {2017})}\BibitemShut {NoStop}%
\bibitem [{\citenamefont {Garc{\'\i}a-Mata}\ \emph {et~al.}(2019)\citenamefont
  {Garc{\'\i}a-Mata}, \citenamefont {Martin}, \citenamefont {Dubertrand},
  \citenamefont {Giraud}, \citenamefont {Georgeot},\ and\ \citenamefont
  {Lemari{\'e}}}]{Lemarie19two_loc_lengths}%
  \BibitemOpen
  \bibfield  {author} {\bibinfo {author} {\bibfnamefont {I}~\bibnamefont
  {Garc{\'\i}a-Mata}}, \bibinfo {author} {\bibfnamefont {J}~\bibnamefont
  {Martin}}, \bibinfo {author} {\bibfnamefont {R}~\bibnamefont {Dubertrand}},
  \bibinfo {author} {\bibfnamefont {O}~\bibnamefont {Giraud}}, \bibinfo
  {author} {\bibfnamefont {B}~\bibnamefont {Georgeot}}, \ and\ \bibinfo
  {author} {\bibfnamefont {G}~\bibnamefont {Lemari{\'e}}},\ }\href@noop {}
  {\enquote {\bibinfo {title} {Two localization lengths in the anderson
  transition on random graphs},}\ } (\bibinfo {year} {2019}),\ \Eprint
  {http://arxiv.org/abs/1904.08869} {arXiv:1904.08869} \BibitemShut {NoStop}%
\bibitem [{\citenamefont {Parisi}\ \emph {et~al.}(2018)\citenamefont {Parisi},
  \citenamefont {Pascazio}, \citenamefont {Pietracaprina}, \citenamefont
  {Ros},\ and\ \citenamefont {Scardicchio}}]{Parisi2018anderson}%
  \BibitemOpen
  \bibfield  {author} {\bibinfo {author} {\bibfnamefont {G.}~\bibnamefont
  {Parisi}}, \bibinfo {author} {\bibfnamefont {S.}~\bibnamefont {Pascazio}},
  \bibinfo {author} {\bibfnamefont {F.}~\bibnamefont {Pietracaprina}}, \bibinfo
  {author} {\bibfnamefont {V.}~\bibnamefont {Ros}}, \ and\ \bibinfo {author}
  {\bibfnamefont {A.}~\bibnamefont {Scardicchio}},\ }\href@noop {} {\enquote
  {\bibinfo {title} {On the anderson transition on the bethe lattice},}\ }
  (\bibinfo {year} {2018}),\ \Eprint {http://arxiv.org/abs/1812.03531}
  {arXiv:1812.03531} \BibitemShut {NoStop}%
\bibitem [{\citenamefont {Savitz}\ \emph {et~al.}(2019)\citenamefont {Savitz},
  \citenamefont {Peng},\ and\ \citenamefont {Refael}}]{Savitz19Wegner_flow}%
  \BibitemOpen
  \bibfield  {author} {\bibinfo {author} {\bibfnamefont {Samuel}\ \bibnamefont
  {Savitz}}, \bibinfo {author} {\bibfnamefont {Changnan}\ \bibnamefont {Peng}},
  \ and\ \bibinfo {author} {\bibfnamefont {Gil}\ \bibnamefont {Refael}},\
  }\bibfield  {title} {\enquote {\bibinfo {title} {Anderson localization on the
  bethe lattice using cages and the wegner flow},}\ }\href {\doibase
  10.1103/PhysRevB.100.094201} {\bibfield  {journal} {\bibinfo  {journal}
  {Phys. Rev. B}\ }\textbf {\bibinfo {volume} {100}},\ \bibinfo {pages}
  {094201} (\bibinfo {year} {2019})}\BibitemShut {NoStop}%
\bibitem [{\citenamefont {Biroli}\ \emph {et~al.}(2012)\citenamefont {Biroli},
  \citenamefont {Ribeiro-Teixeira},\ and\ \citenamefont
  {Tarzia}}]{Biroli:2012vk}%
  \BibitemOpen
  \bibfield  {author} {\bibinfo {author} {\bibfnamefont {G}~\bibnamefont
  {Biroli}}, \bibinfo {author} {\bibfnamefont {A~C}\ \bibnamefont
  {Ribeiro-Teixeira}}, \ and\ \bibinfo {author} {\bibfnamefont {M}~\bibnamefont
  {Tarzia}},\ }\href@noop {} {\enquote {\bibinfo {title} {Difference between
  level statistics, ergodicity and localization transitions on the {Bethe}
  lattice},}\ } (\bibinfo {year} {2012}),\ \Eprint
  {http://arxiv.org/abs/1211.7334} {arXiv:1211.7334} \BibitemShut {NoStop}%
\bibitem [{\citenamefont {Biroli}\ and\ \citenamefont
  {Tarzia}(2018)}]{Biroli2018delocalization}%
  \BibitemOpen
  \bibfield  {author} {\bibinfo {author} {\bibfnamefont {G.}~\bibnamefont
  {Biroli}}\ and\ \bibinfo {author} {\bibfnamefont {M.}~\bibnamefont
  {Tarzia}},\ }\href@noop {} {\enquote {\bibinfo {title} {Delocalization and
  ergodicity of the anderson model on bethe lattices},}\ } (\bibinfo {year}
  {2018}),\ \Eprint {http://arxiv.org/abs/1810.07545} {arXiv:1810.07545}
  \BibitemShut {NoStop}%
\bibitem [{\citenamefont {Metz}\ and\ \citenamefont
  {Perez~Castillo}(2016)}]{Metz16PRL}%
  \BibitemOpen
  \bibfield  {author} {\bibinfo {author} {\bibfnamefont {F.~L.}\ \bibnamefont
  {Metz}}\ and\ \bibinfo {author} {\bibfnamefont {I.}~\bibnamefont
  {Perez~Castillo}},\ }\bibfield  {title} {\enquote {\bibinfo {title} {Large
  deviation function for the number of eigenvalues of sparse random graphs
  inside an interval},}\ }\href {\doibase 10.1103/PhysRevLett.117.104101}
  {\bibfield  {journal} {\bibinfo  {journal} {Phys. Rev. Lett.}\ }\textbf
  {\bibinfo {volume} {117}},\ \bibinfo {pages} {104101} (\bibinfo {year}
  {2016})}\BibitemShut {NoStop}%
\bibitem [{\citenamefont {Metz}\ and\ \citenamefont
  {Castillo}(2017)}]{Metz17PRB}%
  \BibitemOpen
  \bibfield  {author} {\bibinfo {author} {\bibfnamefont {F.~L.}\ \bibnamefont
  {Metz}}\ and\ \bibinfo {author} {\bibfnamefont {I.~P.}\ \bibnamefont
  {Castillo}},\ }\bibfield  {title} {\enquote {\bibinfo {title} {Level
  compressibility for the anderson model on regular random graphs and the
  eigenvalue statistics in the extended phase},}\ }\href {\doibase
  10.1103/PhysRevB.96.064202} {\bibfield  {journal} {\bibinfo  {journal} {Phys.
  Rev. B}\ }\textbf {\bibinfo {volume} {96}},\ \bibinfo {pages} {064202}
  (\bibinfo {year} {2017})}\BibitemShut {NoStop}%
\bibitem [{\citenamefont {Aizenman}\ and\ \citenamefont
  {Warzel}(2011)}]{aizenman2011extended}%
  \BibitemOpen
  \bibfield  {author} {\bibinfo {author} {\bibfnamefont {Michael}\ \bibnamefont
  {Aizenman}}\ and\ \bibinfo {author} {\bibfnamefont {Simone}\ \bibnamefont
  {Warzel}},\ }\bibfield  {title} {\enquote {\bibinfo {title} {Extended states
  in a lifshitz tail regime for random schr\"odinger operators on trees},}\
  }\href {\doibase 10.1103/PhysRevLett.106.136804} {\bibfield  {journal}
  {\bibinfo  {journal} {Phys. Rev. Lett.}\ }\textbf {\bibinfo {volume} {106}},\
  \bibinfo {pages} {136804} (\bibinfo {year} {2011})}\BibitemShut {NoStop}%
\bibitem [{\citenamefont {Aizenman}(2013)}]{Aizenman2013}%
  \BibitemOpen
  \bibfield  {author} {\bibinfo {author} {\bibfnamefont {M.~Warzel~S.}\
  \bibnamefont {Aizenman}},\ }\bibfield  {title} {\enquote {\bibinfo {title}
  {{Resonant delocalization for random Schr{\"o}dinger operators on tree
  graphs}},}\ }\href {\doibase 10.4171/JEMS/389} {\bibfield  {journal}
  {\bibinfo  {journal} {Journal of the European Mathematical Society}\ }\textbf
  {\bibinfo {volume} {15}},\ \bibinfo {pages} {1167--1222} (\bibinfo {year}
  {2013})}\BibitemShut {NoStop}%
\bibitem [{\citenamefont {Bapst}(2014)}]{Bapst2014HighConnectivity}%
  \BibitemOpen
  \bibfield  {author} {\bibinfo {author} {\bibfnamefont {V.}~\bibnamefont
  {Bapst}},\ }\bibfield  {title} {\enquote {\bibinfo {title} {{The large
  connectivity limit of the Anderson model on tree graphs}},}\ }\href {\doibase
  10.1063/1.4894055} {\bibfield  {journal} {\bibinfo  {journal} {Journal of
  Mathematical Physics}\ }\textbf {\bibinfo {volume} {55}},\ \bibinfo {eid}
  {092101} (\bibinfo {year} {2014}),\ 10.1063/1.4894055}\BibitemShut {NoStop}%
\bibitem [{\citenamefont {Mirlin}\ and\ \citenamefont
  {Fyodorov}(1991)}]{mirlin1991localization}%
  \BibitemOpen
  \bibfield  {author} {\bibinfo {author} {\bibfnamefont {A.~D}\ \bibnamefont
  {Mirlin}}\ and\ \bibinfo {author} {\bibfnamefont {Yan~V}\ \bibnamefont
  {Fyodorov}},\ }\bibfield  {title} {\enquote {\bibinfo {title} {{Localization
  transition in the Anderson model on the {Bethe} lattice: spontaneous symmetry
  breaking and correlation functions}},}\ }\href
  {https://www.sciencedirect.com/science/article/pii/055032139190028V}
  {\bibfield  {journal} {\bibinfo  {journal} {Nucl. Phys. B}\ }\textbf
  {\bibinfo {volume} {366}},\ \bibinfo {pages} {507} (\bibinfo {year}
  {1991})}\BibitemShut {NoStop}%
\bibitem [{\citenamefont {Tikhonov}\ \emph {et~al.}(2016)\citenamefont
  {Tikhonov}, \citenamefont {Mirlin},\ and\ \citenamefont {Skvortsov}}]{Tik16}%
  \BibitemOpen
  \bibfield  {author} {\bibinfo {author} {\bibfnamefont {K.~S.}\ \bibnamefont
  {Tikhonov}}, \bibinfo {author} {\bibfnamefont {A.~D.}\ \bibnamefont
  {Mirlin}}, \ and\ \bibinfo {author} {\bibfnamefont {M.~A.}\ \bibnamefont
  {Skvortsov}},\ }\bibfield  {title} {\enquote {\bibinfo {title} {Anderson
  localization and ergodicity on random regular graphs},}\ }\href {\doibase
  10.1103/PhysRevB.94.220203} {\bibfield  {journal} {\bibinfo  {journal} {Phys.
  Rev. B}\ }\textbf {\bibinfo {volume} {94}},\ \bibinfo {pages} {220203(R)}
  (\bibinfo {year} {2016})}\BibitemShut {NoStop}%
\bibitem [{\citenamefont {Tikhonov}\ and\ \citenamefont
  {Mirlin}(2016)}]{TikMir16}%
  \BibitemOpen
  \bibfield  {author} {\bibinfo {author} {\bibfnamefont {K.~S.}\ \bibnamefont
  {Tikhonov}}\ and\ \bibinfo {author} {\bibfnamefont {A.~D.}\ \bibnamefont
  {Mirlin}},\ }\bibfield  {title} {\enquote {\bibinfo {title} {Fractality of
  wave functions on a cayley tree: Difference between tree and locally treelike
  graph without boundary},}\ }\href {\doibase 10.1103/PhysRevB.94.184203}
  {\bibfield  {journal} {\bibinfo  {journal} {Phys. Rev. B}\ }\textbf {\bibinfo
  {volume} {94}},\ \bibinfo {pages} {184203} (\bibinfo {year}
  {2016})}\BibitemShut {NoStop}%
\bibitem [{\citenamefont {Sonner}\ \emph {et~al.}(2017)\citenamefont {Sonner},
  \citenamefont {Tikhonov},\ and\ \citenamefont {Mirlin}}]{Sonner17}%
  \BibitemOpen
  \bibfield  {author} {\bibinfo {author} {\bibfnamefont {M.}~\bibnamefont
  {Sonner}}, \bibinfo {author} {\bibfnamefont {K.~S.}\ \bibnamefont
  {Tikhonov}}, \ and\ \bibinfo {author} {\bibfnamefont {A.~D.}\ \bibnamefont
  {Mirlin}},\ }\bibfield  {title} {\enquote {\bibinfo {title} {Multifractality
  of wave functions on a cayley tree: From root to leaves},}\ }\href {\doibase
  10.1103/PhysRevB.96.214204} {\bibfield  {journal} {\bibinfo  {journal} {Phys.
  Rev. B}\ }\textbf {\bibinfo {volume} {96}},\ \bibinfo {pages} {214204}
  (\bibinfo {year} {2017})}\BibitemShut {NoStop}%
\bibitem [{\citenamefont {Tikhonov}\ and\ \citenamefont
  {Mirlin}(2019{\natexlab{b}})}]{Tikh2019_K(w)}%
  \BibitemOpen
  \bibfield  {author} {\bibinfo {author} {\bibfnamefont {K.~S.}\ \bibnamefont
  {Tikhonov}}\ and\ \bibinfo {author} {\bibfnamefont {A.~D.}\ \bibnamefont
  {Mirlin}},\ }\bibfield  {title} {\enquote {\bibinfo {title} {Statistics of
  eigenstates near the localization transition on random regular graphs},}\
  }\href {\doibase 10.1103/PhysRevB.99.024202} {\bibfield  {journal} {\bibinfo
  {journal} {Phys. Rev. B}\ }\textbf {\bibinfo {volume} {99}},\ \bibinfo
  {pages} {024202} (\bibinfo {year} {2019}{\natexlab{b}})}\BibitemShut
  {NoStop}%
\bibitem [{\citenamefont {Srednicki}(1994)}]{Sred94}%
  \BibitemOpen
  \bibfield  {author} {\bibinfo {author} {\bibfnamefont {M.}~\bibnamefont
  {Srednicki}},\ }\bibfield  {title} {\enquote {\bibinfo {title} {Chaos and
  quantum thermalization},}\ }\href {\doibase 10.1103/PhysRevE.50.888}
  {\bibfield  {journal} {\bibinfo  {journal} {Phys. Rev. E}\ }\textbf {\bibinfo
  {volume} {50}},\ \bibinfo {pages} {888--901} (\bibinfo {year}
  {1994})}\BibitemShut {NoStop}%
\bibitem [{\citenamefont {Deutsch}(1991)}]{Deut91}%
  \BibitemOpen
  \bibfield  {author} {\bibinfo {author} {\bibfnamefont {J.~M.}\ \bibnamefont
  {Deutsch}},\ }\bibfield  {title} {\enquote {\bibinfo {title} {Quantum
  statistical mechanics in a closed system},}\ }\href {\doibase
  10.1103/PhysRevA.43.2046} {\bibfield  {journal} {\bibinfo  {journal} {Phys.
  Rev. A}\ }\textbf {\bibinfo {volume} {43}},\ \bibinfo {pages} {2046--2049}
  (\bibinfo {year} {1991})}\BibitemShut {NoStop}%
\bibitem [{\citenamefont {Rigol}\ and\ \citenamefont
  {Srednicki}(2012)}]{Rigol2012}%
  \BibitemOpen
  \bibfield  {author} {\bibinfo {author} {\bibfnamefont {M.}~\bibnamefont
  {Rigol}}\ and\ \bibinfo {author} {\bibfnamefont {M.}~\bibnamefont
  {Srednicki}},\ }\bibfield  {title} {\enquote {\bibinfo {title} {Alternatives
  to eigenstate thermalization},}\ }\href {\doibase
  10.1103/PhysRevLett.108.110601} {\bibfield  {journal} {\bibinfo  {journal}
  {Phys. Rev. Lett.}\ }\textbf {\bibinfo {volume} {108}},\ \bibinfo {pages}
  {110601} (\bibinfo {year} {2012})}\BibitemShut {NoStop}%
\bibitem [{\citenamefont {{Smith}}\ \emph {et~al.}(2016)\citenamefont
  {{Smith}}, \citenamefont {{Lee}}, \citenamefont {{Richerme}}, \citenamefont
  {{Neyenhuis}}, \citenamefont {{Hess}}, \citenamefont {{Hauke}}, \citenamefont
  {{Heyl}}, \citenamefont {{Huse}},\ and\ \citenamefont
  {{Monroe}}}]{monroe2015}%
  \BibitemOpen
  \bibfield  {author} {\bibinfo {author} {\bibfnamefont {J.}~\bibnamefont
  {{Smith}}}, \bibinfo {author} {\bibfnamefont {A.}~\bibnamefont {{Lee}}},
  \bibinfo {author} {\bibfnamefont {P.}~\bibnamefont {{Richerme}}}, \bibinfo
  {author} {\bibfnamefont {B.}~\bibnamefont {{Neyenhuis}}}, \bibinfo {author}
  {\bibfnamefont {P.~W.}\ \bibnamefont {{Hess}}}, \bibinfo {author}
  {\bibfnamefont {P.}~\bibnamefont {{Hauke}}}, \bibinfo {author} {\bibfnamefont
  {M.}~\bibnamefont {{Heyl}}}, \bibinfo {author} {\bibfnamefont {D.~A.}\
  \bibnamefont {{Huse}}}, \ and\ \bibinfo {author} {\bibfnamefont
  {C.}~\bibnamefont {{Monroe}}},\ }\bibfield  {title} {\enquote {\bibinfo
  {title} {{Many-body localization in a quantum simulator with programmable
  random disorder}},}\ }\href
  {http://www.nature.com/nphys/journal/v12/n10/full/nphys3783.html} {\bibfield
  {journal} {\bibinfo  {journal} {Nat. Phys.}\ }\textbf {\bibinfo {volume}
  {12}},\ \bibinfo {pages} {907--911} (\bibinfo {year} {2016})}\BibitemShut
  {NoStop}%
\bibitem [{\citenamefont {Hauke}\ and\ \citenamefont {Heyl}(2015)}]{Hauke15}%
  \BibitemOpen
  \bibfield  {author} {\bibinfo {author} {\bibfnamefont {P.}~\bibnamefont
  {Hauke}}\ and\ \bibinfo {author} {\bibfnamefont {M.}~\bibnamefont {Heyl}},\
  }\bibfield  {title} {\enquote {\bibinfo {title} {Many-body localization and
  quantum ergodicity in disordered long-range ising models},}\ }\href {\doibase
  10.1103/PhysRevB.92.134204} {\bibfield  {journal} {\bibinfo  {journal} {Phys.
  Rev. B}\ }\textbf {\bibinfo {volume} {92}},\ \bibinfo {pages} {134204}
  (\bibinfo {year} {2015})}\BibitemShut {NoStop}%
\bibitem [{\citenamefont {Ketzmerick}\ \emph {et~al.}(1997)\citenamefont
  {Ketzmerick}, \citenamefont {Kruse}, \citenamefont {Kraut},\ and\
  \citenamefont {Geisel}}]{Ketzmerick1997}%
  \BibitemOpen
  \bibfield  {author} {\bibinfo {author} {\bibfnamefont {R.}~\bibnamefont
  {Ketzmerick}}, \bibinfo {author} {\bibfnamefont {K.}~\bibnamefont {Kruse}},
  \bibinfo {author} {\bibfnamefont {S.}~\bibnamefont {Kraut}}, \ and\ \bibinfo
  {author} {\bibfnamefont {T.}~\bibnamefont {Geisel}},\ }\bibfield  {title}
  {\enquote {\bibinfo {title} {{What Determines the Spreading of a Wave
  Packet?}}}\ }\href {\doibase 10.1103/PhysRevLett.79.1959} {\bibfield
  {journal} {\bibinfo  {journal} {Phys. Rev. Lett.}\ }\textbf {\bibinfo
  {volume} {79}},\ \bibinfo {pages} {1959--1963} (\bibinfo {year}
  {1997})}\BibitemShut {NoStop}%
\bibitem [{\citenamefont {Ohtsuki}\ and\ \citenamefont
  {Kawarabayashi}(1997)}]{Ohtsuki1997}%
  \BibitemOpen
  \bibfield  {author} {\bibinfo {author} {\bibfnamefont {Tomi}\ \bibnamefont
  {Ohtsuki}}\ and\ \bibinfo {author} {\bibfnamefont {Tohru}\ \bibnamefont
  {Kawarabayashi}},\ }\bibfield  {title} {\enquote {\bibinfo {title}
  {{Anomalous Diffusion at the Anderson Transitions}},}\ }\href {\doibase
  10.1143/JPSJ.66.314} {\bibfield  {journal} {\bibinfo  {journal} {J. Phys.
  Soc. Japan}\ }\textbf {\bibinfo {volume} {66}},\ \bibinfo {pages} {314--317}
  (\bibinfo {year} {1997})}\BibitemShut {NoStop}%
\bibitem [{Note1()}]{Note1}%
  \BibitemOpen
  \bibinfo {note} {Notice the correct normalization $\DOTSB \sum@ \slimits@ _x
  \Pi (x,t)= 1$, and $\Pi (x,t)\ge 0$.}\BibitemShut {Stop}%
\bibitem [{\citenamefont {Ros}\ \emph {et~al.}(2015)\citenamefont {Ros},
  \citenamefont {M{\"u}ller},\ and\ \citenamefont
  {Scardicchio}}]{ros2015integrals}%
  \BibitemOpen
  \bibfield  {author} {\bibinfo {author} {\bibfnamefont {V}~\bibnamefont
  {Ros}}, \bibinfo {author} {\bibfnamefont {M}~\bibnamefont {M{\"u}ller}}, \
  and\ \bibinfo {author} {\bibfnamefont {A}~\bibnamefont {Scardicchio}},\
  }\bibfield  {title} {\enquote {\bibinfo {title} {{Integrals of motion in the
  many-body localized phase}},}\ }\href
  {http://www.sciencedirect.com/science/article/pii/S0550321314003836}
  {\bibfield  {journal} {\bibinfo  {journal} {Nucl. Phys. B}\ }\textbf
  {\bibinfo {volume} {891}},\ \bibinfo {pages} {420--465} (\bibinfo {year}
  {2015})}\BibitemShut {NoStop}%
\bibitem [{Note2()}]{Note2}%
  \BibitemOpen
  \bibinfo {note} {The single-particle mobility edges have been proven to exist
  at $W\gtrsim K$~\cite {aizenman2011extended}}\BibitemShut {NoStop}%
\bibitem [{\citenamefont {Mehta.}(2004)}]{Mehta}%
  \BibitemOpen
  \bibfield  {author} {\bibinfo {author} {\bibfnamefont {M.L.}\ \bibnamefont
  {Mehta.}},\ }\href@noop {} {\emph {\bibinfo {title} {{Random Matrices,
  $3^{rd}$ ed.}}}}\ (\bibinfo  {publisher} {Elsevier Inc., Amsterdam},\
  \bibinfo {year} {2004})\BibitemShut {NoStop}%
\bibitem [{\citenamefont {Bera}\ \emph {et~al.}(2018)\citenamefont {Bera},
  \citenamefont {De~Tomasi}, \citenamefont {Khaymovich},\ and\ \citenamefont
  {Scardicchio}}]{bera19}%
  \BibitemOpen
  \bibfield  {author} {\bibinfo {author} {\bibfnamefont {S.}~\bibnamefont
  {Bera}}, \bibinfo {author} {\bibfnamefont {G.}~\bibnamefont {De~Tomasi}},
  \bibinfo {author} {\bibfnamefont {I.~M.}\ \bibnamefont {Khaymovich}}, \ and\
  \bibinfo {author} {\bibfnamefont {A.}~\bibnamefont {Scardicchio}},\
  }\bibfield  {title} {\enquote {\bibinfo {title} {Return probability for the
  anderson model on the random regular graph},}\ }\href {\doibase
  10.1103/PhysRevB.98.134205} {\bibfield  {journal} {\bibinfo  {journal} {Phys.
  Rev. B}\ }\textbf {\bibinfo {volume} {98}},\ \bibinfo {pages} {134205}
  (\bibinfo {year} {2018})}\BibitemShut {NoStop}%
\bibitem [{\citenamefont {De~Tomasi}\ \emph
  {et~al.}(2019{\natexlab{b}})\citenamefont {De~Tomasi}, \citenamefont {Amini},
  \citenamefont {Bera}, \citenamefont {Khaymovich},\ and\ \citenamefont
  {Kravtsov}}]{gdt19}%
  \BibitemOpen
  \bibfield  {author} {\bibinfo {author} {\bibfnamefont {G.}~\bibnamefont
  {De~Tomasi}}, \bibinfo {author} {\bibfnamefont {M.}~\bibnamefont {Amini}},
  \bibinfo {author} {\bibfnamefont {S.}~\bibnamefont {Bera}}, \bibinfo {author}
  {\bibfnamefont {I.~M.}\ \bibnamefont {Khaymovich}}, \ and\ \bibinfo {author}
  {\bibfnamefont {V.~E.}\ \bibnamefont {Kravtsov}},\ }\bibfield  {title}
  {\enquote {\bibinfo {title} {{Survival probability in Generalized
  Rosenzweig-Porter random matrix ensemble}},}\ }\href {\doibase
  10.21468/SciPostPhys.6.1.014} {\bibfield  {journal} {\bibinfo  {journal}
  {Scipost Phys.}\ }\textbf {\bibinfo {volume} {6}},\ \bibinfo {pages} {14}
  (\bibinfo {year} {2019}{\natexlab{b}})}\BibitemShut {NoStop}%
\bibitem [{FE-()}]{FE-footnote}%
  \BibitemOpen
  \href@noop {} {}\bibinfo {note} {We define a fully ergodic phase as the phase
  in which the system can be described by the random matrix
  theory.}\BibitemShut {Stop}%
\bibitem [{W3-()}]{W3-footnote}%
  \BibitemOpen
  \href@noop {} {}\bibinfo {note} {In Appendix~\ref{SM:Sec2_Weak_disorder} we
  explicitly show diffusive behavior in this disorder range $W<0.16
  \WAT$.}\BibitemShut {Stop}%
\bibitem [{\citenamefont {Giacometti}(1995)}]{Giacometti95}%
  \BibitemOpen
  \bibfield  {author} {\bibinfo {author} {\bibfnamefont {A.}~\bibnamefont
  {Giacometti}},\ }\bibfield  {title} {\enquote {\bibinfo {title} {Exact closed
  form of the return probability on the bethe lattice},}\ }\href {\doibase
  10.1088/0305-4470/28/1/003} {\bibfield  {journal} {\bibinfo  {journal} {J.
  Phys. A}\ }\textbf {\bibinfo {volume} {28}},\ \bibinfo {pages} {L13--L17}
  (\bibinfo {year} {1995})}\BibitemShut {NoStop}%
\bibitem [{Note3()}]{Note3}%
  \BibitemOpen
  \bibinfo {note} {For $W<0.16 W_{\protect \text {AT}}$ the propagation is
  diffusive with $\beta =1$ and consistent with the fully-ergodic behavior
  given by Eq.~\protect \textup {\hbox {\mathsurround \z@ \protect \normalfont
  (\ignorespaces \ref {eq:Return}\unskip \@@italiccorr )}}.}\BibitemShut
  {Stop}%
\bibitem [{Note4()}]{Note4}%
  \BibitemOpen
  \bibinfo {note} {$X_{\protect \text {front}}(t)$ it is given by the value of
  $x$ for which $\Pi (x,t)$ has a maximum}\BibitemShut {NoStop}%
\bibitem [{\citenamefont {Edwards}\ and\ \citenamefont
  {Thouless}(1972)}]{Edwards_1972}%
  \BibitemOpen
  \bibfield  {author} {\bibinfo {author} {\bibfnamefont {J~T}\ \bibnamefont
  {Edwards}}\ and\ \bibinfo {author} {\bibfnamefont {D~J}\ \bibnamefont
  {Thouless}},\ }\bibfield  {title} {\enquote {\bibinfo {title} {Numerical
  studies of localization in disordered systems},}\ }\href {\doibase
  10.1088/0022-3719/5/8/007} {\bibfield  {journal} {\bibinfo  {journal}
  {Journal of Physics C: Solid State Physics}\ }\textbf {\bibinfo {volume}
  {5}},\ \bibinfo {pages} {807--820} (\bibinfo {year} {1972})}\BibitemShut
  {NoStop}%
\bibitem [{Note5()}]{Note5}%
  \BibitemOpen
  \bibinfo {note} {Note that both collapses, Eqs.~\protect \textup {\hbox
  {\mathsurround \z@ \protect \normalfont (\ignorespaces \ref
  {eq:front_X_f-x}\unskip \@@italiccorr )}} and \protect \textup {\hbox
  {\mathsurround \z@ \protect \normalfont (\ignorespaces \ref
  {eq:Factorization}\unskip \@@italiccorr )}}, are consistent to each other
  provided the exponential behavior of $f(z)$ in Eq.~\protect \textup {\hbox
  {\mathsurround \z@ \protect \normalfont (\ignorespaces \ref
  {eq:front_X_f-x}\unskip \@@italiccorr )}}, with the maximal value $\Pi
  (X_{\protect \text {front}}(t),t)$ given by $g(X_{\protect \text
  {front}})\left [\Pi (0,t)- \Pi (0,\infty )\right ]$ due to the normalization
  condition of $\Pi (x,t)$ (see Appendix~\ref {SM:Sec4_Collapse} for more
  details).}\BibitemShut {Stop}%
\bibitem [{\citenamefont {Serbyn}\ \emph {et~al.}(2017)\citenamefont {Serbyn},
  \citenamefont {Papi\ifmmode~\acute{c}\else \'{c}\fi{}},\ and\ \citenamefont
  {Abanin}}]{Serb_th17}%
  \BibitemOpen
  \bibfield  {author} {\bibinfo {author} {\bibfnamefont {M.}~\bibnamefont
  {Serbyn}}, \bibinfo {author} {\bibfnamefont {Z.}~\bibnamefont
  {Papi\ifmmode~\acute{c}\else \'{c}\fi{}}}, \ and\ \bibinfo {author}
  {\bibfnamefont {D.~A.}\ \bibnamefont {Abanin}},\ }\bibfield  {title}
  {\enquote {\bibinfo {title} {Thouless energy and multifractality across the
  many-body localization transition},}\ }\href {\doibase
  10.1103/PhysRevB.96.104201} {\bibfield  {journal} {\bibinfo  {journal} {Phys.
  Rev. B}\ }\textbf {\bibinfo {volume} {96}},\ \bibinfo {pages} {104201}
  (\bibinfo {year} {2017})}\BibitemShut {NoStop}%
\bibitem [{\citenamefont {Bertrand}\ and\ \citenamefont
  {Garc\'{\i}a-Garc\'{\i}a}(2016)}]{Gar_th}%
  \BibitemOpen
  \bibfield  {author} {\bibinfo {author} {\bibfnamefont {C.~L.}\ \bibnamefont
  {Bertrand}}\ and\ \bibinfo {author} {\bibfnamefont {A.~M.}\ \bibnamefont
  {Garc\'{\i}a-Garc\'{\i}a}},\ }\bibfield  {title} {\enquote {\bibinfo {title}
  {Anomalous thouless energy and critical statistics on the metallic side of
  the many-body localization transition},}\ }\href {\doibase
  10.1103/PhysRevB.94.144201} {\bibfield  {journal} {\bibinfo  {journal} {Phys.
  Rev. B}\ }\textbf {\bibinfo {volume} {94}},\ \bibinfo {pages} {144201}
  (\bibinfo {year} {2016})}\BibitemShut {NoStop}%
\bibitem [{Note6()}]{Note6}%
  \BibitemOpen
  \bibinfo {note} {{\protect \color {black}In the literature~\cite
  {BogomolnyPLRBM2018,Nosov2019correlation,nosov2019mixtures,Baecker2019} this
  kind of phase is usually called {\protect \it weakly} ergodic when the
  eigenfunctions occupy a finite, but small fraction $f_I\ll 1$ of the Hilbert
  space, meaning that the system splits into a finite number $\sim 1/f_I$ of
  nearly independent sectors which are ergodic within themselves, but not
  between each other.}}\BibitemShut {Stop}%
\bibitem [{\citenamefont {Aizenman}\ and\ \citenamefont
  {Warzel}(2012)}]{Aizenman2012absolutely}%
  \BibitemOpen
  \bibfield  {author} {\bibinfo {author} {\bibfnamefont {M.}~\bibnamefont
  {Aizenman}}\ and\ \bibinfo {author} {\bibfnamefont {S.}~\bibnamefont
  {Warzel}},\ }\bibfield  {title} {\enquote {\bibinfo {title} {Absolutely
  continuous spectrum implies ballistic transport for quantum particles in a
  random potential on tree graphs},}\ }\href
  {https://doi.org/10.1063/1.4714617} {\bibfield  {journal} {\bibinfo
  {journal} {Journal of Mathematical Physics}\ }\textbf {\bibinfo {volume}
  {53}},\ \bibinfo {pages} {095205} (\bibinfo {year} {2012})}\BibitemShut
  {NoStop}%
\bibitem [{\citenamefont {Luitz}\ \emph {et~al.}(2019)\citenamefont {Luitz},
  \citenamefont {Khaymovich},\ and\ \citenamefont
  {Lev}}]{luitz2019multifractality}%
  \BibitemOpen
  \bibfield  {author} {\bibinfo {author} {\bibfnamefont {David~J}\ \bibnamefont
  {Luitz}}, \bibinfo {author} {\bibfnamefont {Ivan}\ \bibnamefont
  {Khaymovich}}, \ and\ \bibinfo {author} {\bibfnamefont {Yevgeny~Bar}\
  \bibnamefont {Lev}},\ }\href@noop {} {\enquote {\bibinfo {title}
  {Multifractality and its role in anomalous transport in the disordered xxz
  spin-chain},}\ } (\bibinfo {year} {2019}),\ \Eprint
  {http://arxiv.org/abs/1909.06380} {arXiv:1909.06380} \BibitemShut {NoStop}%
\bibitem [{\citenamefont {Bogomolny}\ and\ \citenamefont
  {Sieber}(2018)}]{BogomolnyPLRBM2018}%
  \BibitemOpen
  \bibfield  {author} {\bibinfo {author} {\bibfnamefont {E.}~\bibnamefont
  {Bogomolny}}\ and\ \bibinfo {author} {\bibfnamefont {M.}~\bibnamefont
  {Sieber}},\ }\bibfield  {title} {\enquote {\bibinfo {title} {Power-law random
  banded matrices and ultrametric matrices: Eigenvector distribution in the
  intermediate regime},}\ }\href {\doibase 10.1103/PhysRevE.98.042116}
  {\bibfield  {journal} {\bibinfo  {journal} {Phys. Rev. E}\ }\textbf {\bibinfo
  {volume} {98}},\ \bibinfo {pages} {042116} (\bibinfo {year}
  {2018})}\BibitemShut {NoStop}%
\bibitem [{\citenamefont {Nosov}\ \emph {et~al.}(2019)\citenamefont {Nosov},
  \citenamefont {Khaymovich},\ and\ \citenamefont
  {Kravtsov}}]{Nosov2019correlation}%
  \BibitemOpen
  \bibfield  {author} {\bibinfo {author} {\bibfnamefont {P.~A.}\ \bibnamefont
  {Nosov}}, \bibinfo {author} {\bibfnamefont {I.~M.}\ \bibnamefont
  {Khaymovich}}, \ and\ \bibinfo {author} {\bibfnamefont {V.~E.}\ \bibnamefont
  {Kravtsov}},\ }\bibfield  {title} {\enquote {\bibinfo {title}
  {Correlation-induced localization},}\ }\href
  {https://doi.org/10.1103/PhysRevB.99.104203} {\bibfield  {journal} {\bibinfo
  {journal} {Physical Review B}\ }\textbf {\bibinfo {volume} {99}},\ \bibinfo
  {pages} {104203} (\bibinfo {year} {2019})}\BibitemShut {NoStop}%
\bibitem [{\citenamefont {Nosov}\ and\ \citenamefont
  {Khaymovich}(2019)}]{nosov2019mixtures}%
  \BibitemOpen
  \bibfield  {author} {\bibinfo {author} {\bibfnamefont {P.~A.}\ \bibnamefont
  {Nosov}}\ and\ \bibinfo {author} {\bibfnamefont {I.~M.}\ \bibnamefont
  {Khaymovich}},\ }\bibfield  {title} {\enquote {\bibinfo {title} {Robustness
  of delocalization to the inclusion of soft constraints in long-range random
  models},}\ }\href {\doibase 10.1103/PhysRevB.99.224208} {\bibfield  {journal}
  {\bibinfo  {journal} {Phys. Rev. B}\ }\textbf {\bibinfo {volume} {99}},\
  \bibinfo {pages} {224208} (\bibinfo {year} {2019})}\BibitemShut {NoStop}%
\bibitem [{\citenamefont {B{\"{a}}cker}\ \emph {et~al.}(2019)\citenamefont
  {B{\"{a}}cker}, \citenamefont {Haque},\ and\ \citenamefont
  {Khaymovich}}]{Baecker2019}%
  \BibitemOpen
  \bibfield  {author} {\bibinfo {author} {\bibfnamefont {Arnd}\ \bibnamefont
  {B{\"{a}}cker}}, \bibinfo {author} {\bibfnamefont {Masudul}\ \bibnamefont
  {Haque}}, \ and\ \bibinfo {author} {\bibfnamefont {Ivan~M}\ \bibnamefont
  {Khaymovich}},\ }\bibfield  {title} {\enquote {\bibinfo {title}
  {{Multifractal dimensions for random matrices, chaotic quantum maps, and
  many-body systems}},}\ }\href {\doibase 10.1103/PhysRevE.100.032117}
  {\bibfield  {journal} {\bibinfo  {journal} {Phys. Rev. E}\ }\textbf {\bibinfo
  {volume} {100}},\ \bibinfo {pages} {032117} (\bibinfo {year}
  {2019})}\BibitemShut {NoStop}%
\end{thebibliography}%

\appendix
\section{Initial wave-packet and dependence on $f$}\label{SM:Sec1_finite-size+f}

As we already discussed in the main text, the usage of the projector $\hat P_{\Delta E}$ spreads slightly the initial delta-function like state $| x_0\rangle$.
In this Supplemental Note we provide evidence that our results are barely affected by the size of the energy shell $\delta E =  f E_{\text{BW}}$ taken.

\begin{figure}[t!]
 \includegraphics[width=0.9\columnwidth]{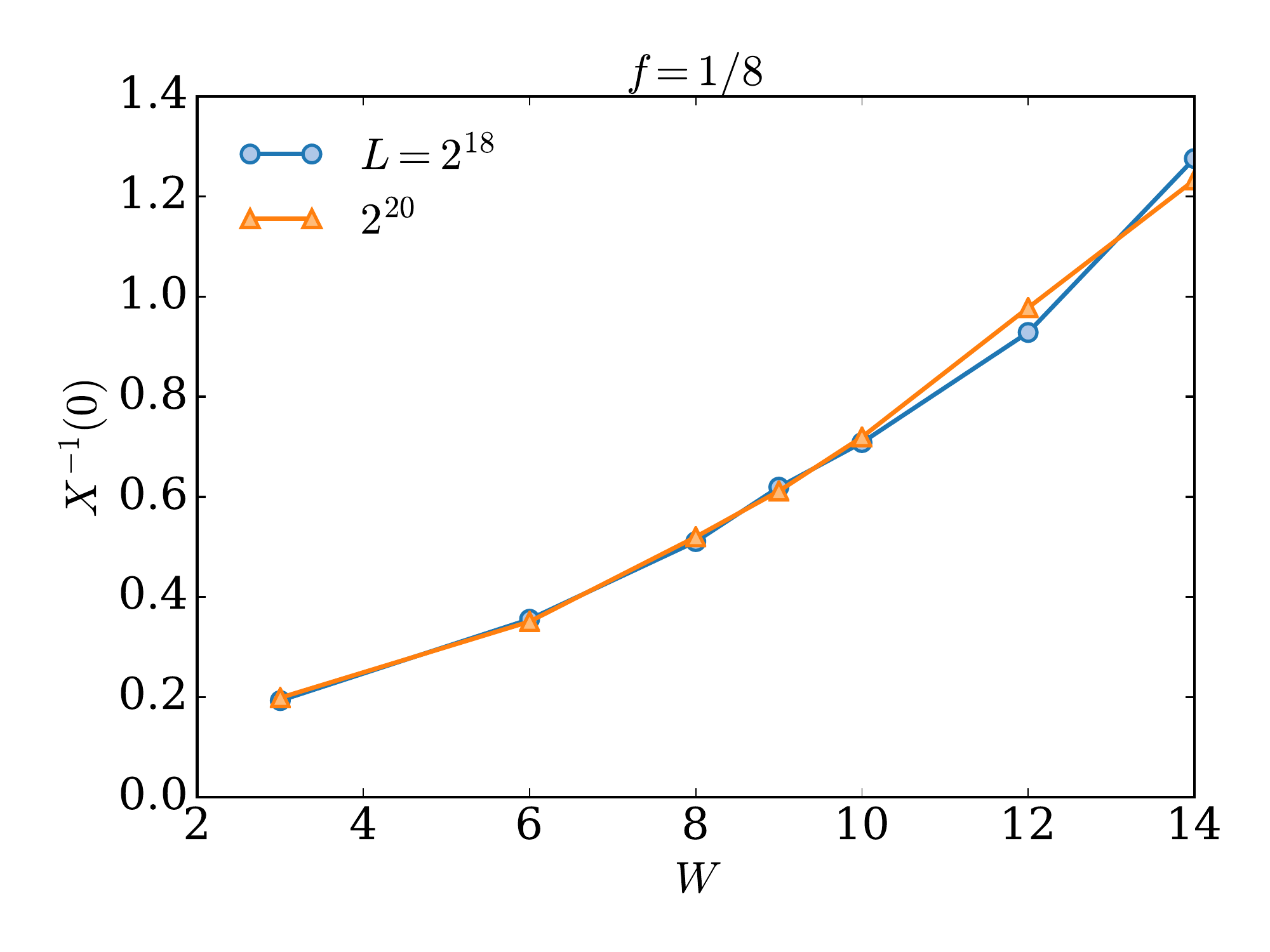}
 \caption{The inverse initial size of the wavepacket versus disorder strength for two largest available system sizes $L=2^{18}$ and $L=2^{20}$.}
\label{fig:S6}
\end{figure}

Figure~\ref{fig:S6} shows the initial size of the wavepacket estimated by $X(0) = \sum_x x\Pxt$ as function of $W$. As expected $X(0)$ is a decreasing function of $W$ and importantly it does not depend on the system size $L$ ($L\gg X(0)$) as shown in  Fig.~\ref{fig:S6}.
The latter confirms that the initial localization of the wavepacket is not affected by the introduction of the projector $\hat P_{\Delta E}$.

\begin{figure}[t!]
 \includegraphics[width=0.9\columnwidth]{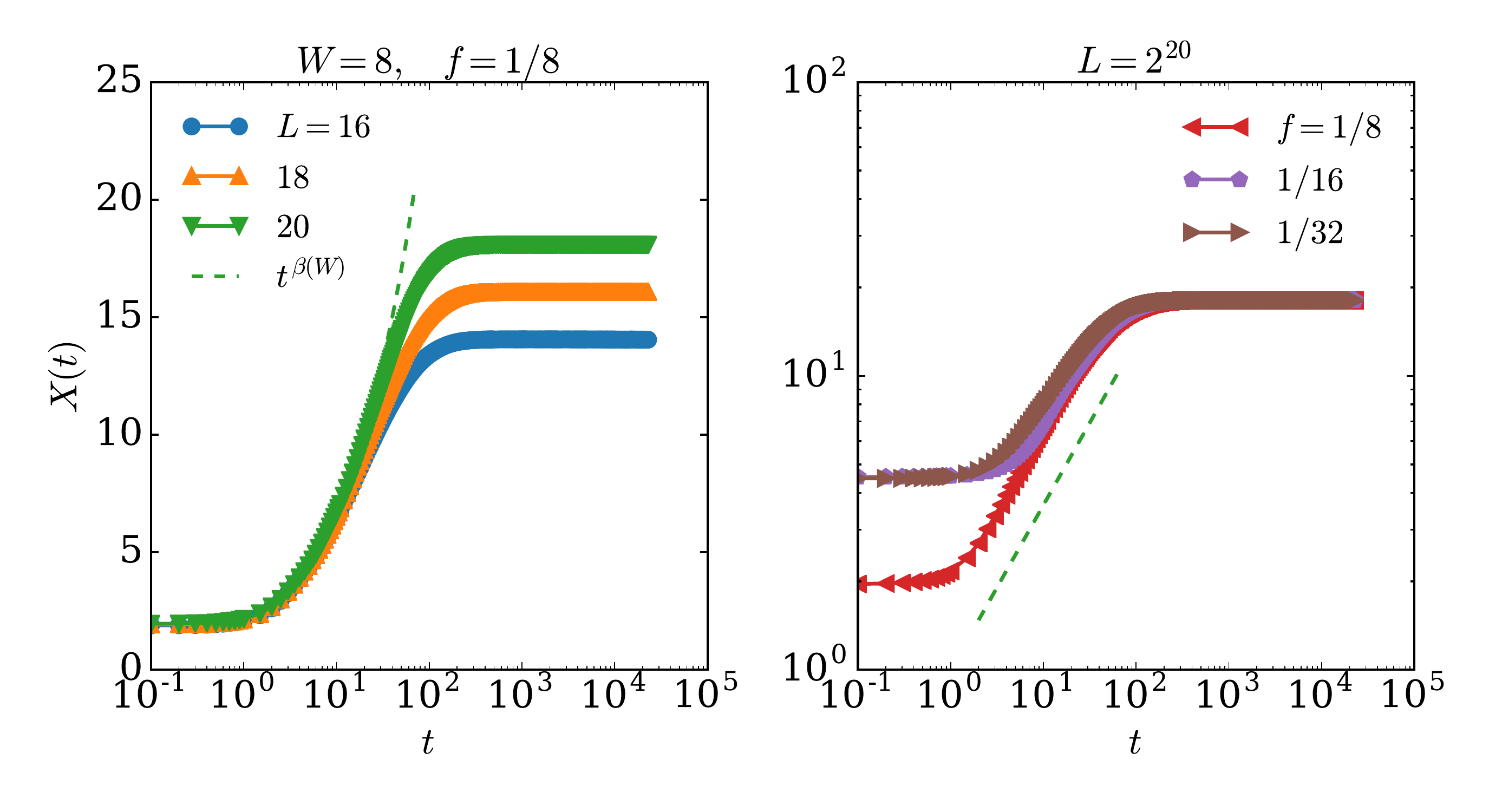}
 \caption{Left panel: Wavepacket size $X(t)$ as a function of time $t$ at $W=8$ and several system size $L\in \{2^{16}, 2^{18}, 2^{20}\}$. The dashed line is a guide for a eyes $X(t) \sim t^{\beta(W)}$ with $\beta(W)=1-W/\WAT$.
 Right panel: The same data $X(t)$ versus $t$ at $W=8$ at fixed $L=2^{20}$ for several energy shell fractions $f$ ($\delta E = f E_{\text{BW}}$).
         }
\label{fig:S7}
\end{figure}

Next, we consider the effects of finite system size $L$ and energy shell $\delta E$ on finite-time dynamics of the wavepacket size $X(t)$.
Left panel of Fig.~\ref{fig:S7} shows $X(t)$ at $W=8$ as a function of time at several system sizes $L\in \{2^{16}, 2^{18}, 2^{20}\}$.
The curves lie on top of each other in the time interval spreading to the finite-size saturation point and enlarging with $L$.
Thus, we do not report any significant finite-size effects at such times.

The effect of the finite energy shell $\delta E = f E_{\text{BW}}$ is shown in the right panel of Fig.~\ref{fig:S7}.
Plots of $X(t)$ versus $t$ for fixed $L=2^{20}$ demonstrate the same behavior in time 
and are barely affected by $f$ in the sub-diffusive regime.
Thus, one should conclude that for the range of energy shells considered there is no change in the dynamics and the results are robust to the value of $f$.

\section{Weak disorder}\label{SM:Sec2_Weak_disorder}

In this Supplemental Note, we focus on the Anderson model $\hat{H}$ on the random regular graph (RRG) at weak disorder $W\lesssim 0.16 \WAT$.
We start our investigation from the inspection of the spreading of the initially localized wave-packet $\hat{P}_{\Delta E}|x_0\rangle$ as we did in the main text. In particular, we give numerical indication that the dynamics of this wavepacket spreading is diffusive.

At weak disorder a fully-ergodic phase has been reported. 
Indeed, at weak disorder not just the IPR scales to zero as the inverse volume of the RRG but also the finite-time dynamics can be described using random matrix theory. 
As a result the return probability has the following form
\begin{equation}
\label{eq:return_erg}
 \Rt=\left (\frac{\sin{\delta E t}}{\delta E  t}\right )^2.
\end{equation}
The aforementioned behavior contrasts to the decay of $\Rt$ at intermediate disorder strength ($\sim e^{-\Gamma t^{\beta(W)}}$), that we discussed in the main text. 

\begin{figure}[h!]
 \includegraphics[width=0.9\columnwidth]{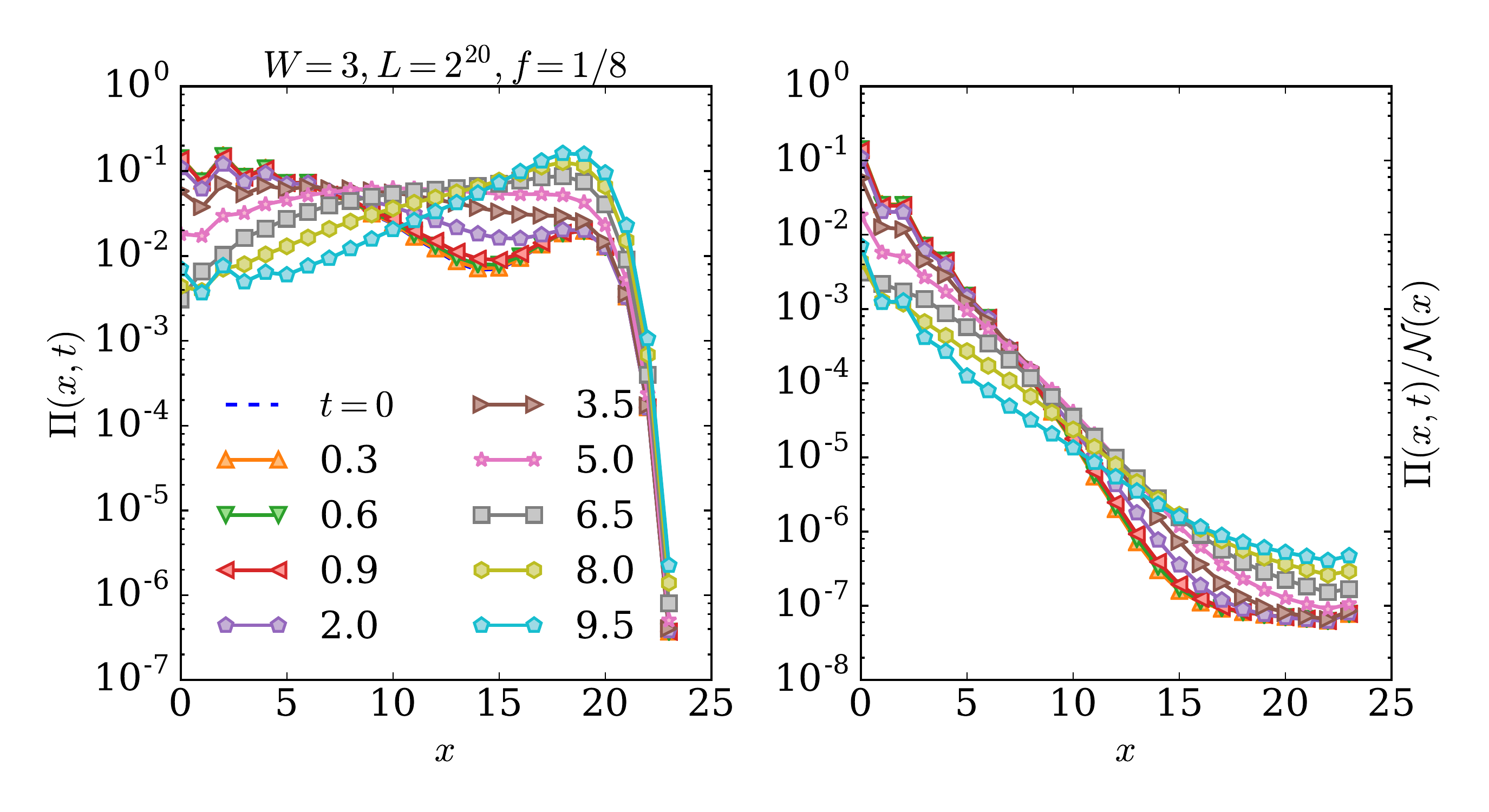}
 \caption{Left panel: $\Pxt$ at weak-disorder $W=3$ as a function of the distance $x$ at several times.
 Right panel: $\Pxt$ renormalized
by the mean number of sites $\N(x)$  at distance $x$ from an initial site state $|x_0 \rangle$.}
 \label{fig:S1}
 \end{figure}

As well as for the case at intermediate disorder strengths, here the relaxation of $\Pxt$ can be also described by the formation of a semi-classical wave-front.
Indeed, in Fig.~\ref{fig:S1} we show the relaxation of the probability distribution $\Pxt$ with the formation of the wave-front $\Xf$ which reaches the diameter of the graph, $\Pi(x,\tTh)\approx \text{const}$, at the time $\tTh(W=3)\approx 5$. written in units of an inverse hopping constant.
As in the main text, this time is natural to call the Thouless time $\tTh$
as it estimates the time scale needed to a wavepacket to spread up to the graph diameter.
As one expects the value of $\tTh$ for $W=3$ is smaller than the one in the case of larger disorder $W=8$ considered in the main text.
\begin{figure}[tb]
 \includegraphics[width=0.9\columnwidth]{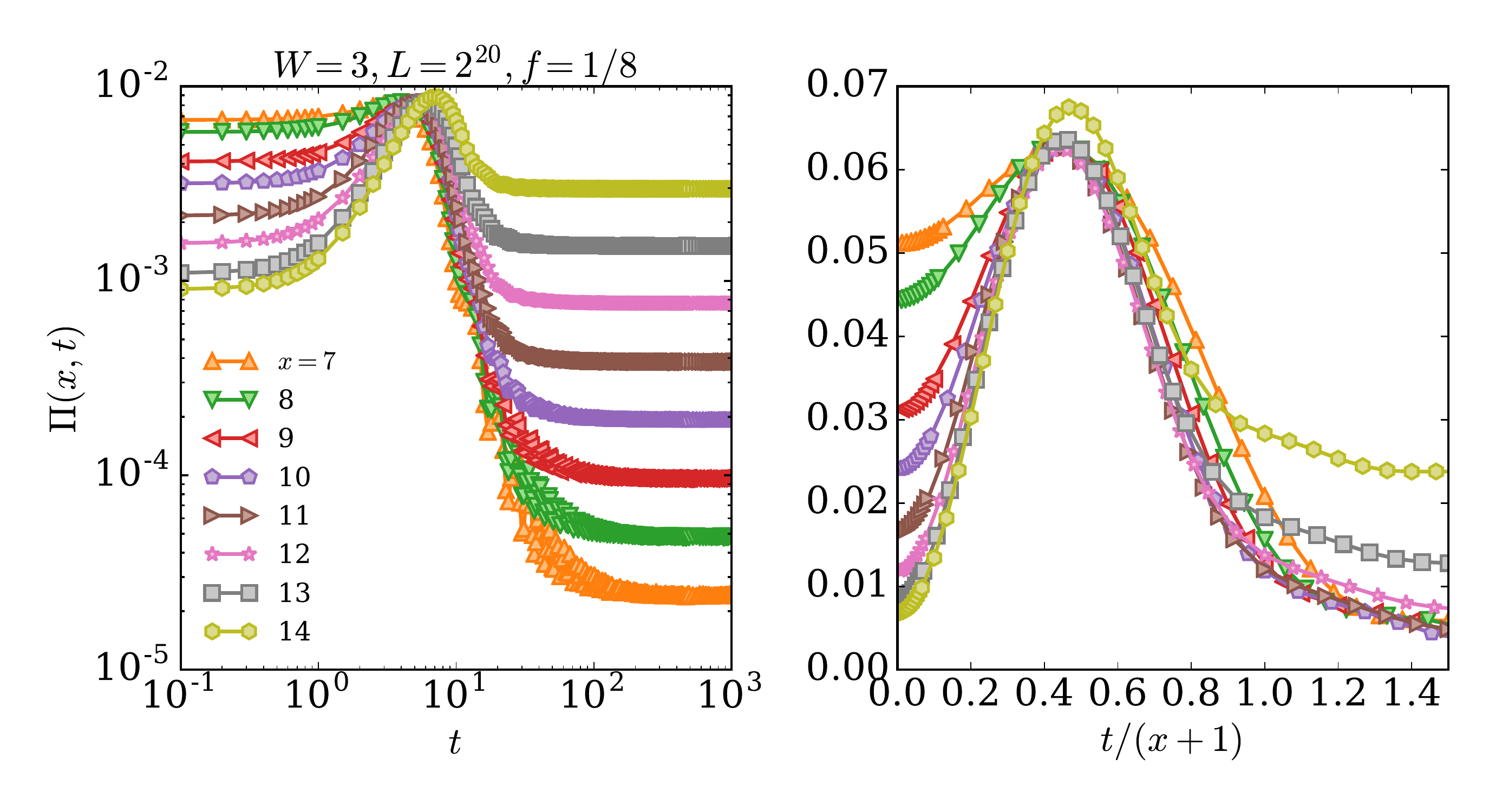}
 \caption{Left panel: $\Pxt$ at $W=3$ as a function of time for several $x>X(0)$.
 Right panel: Ballistic propagation of the wave-front at weak disorder $W=3$. The axis is properly renormalized to show $\Xf\sim t$.}
 \label{fig:S2}
 \end{figure}

More quantitatively the propagation of $\Pxt$ can be characterized by the position of the wave-front in space-time $(x,t)$.
Figure~\ref{fig:S2} shows $\Pxt$ at distances $x$ larger than the initial localization length $\sim X(0)$ of the wave-packet.
At short time (large distance), $x>\Xf$, $\Pxt$ is frozen $\Pxt\approx \PxO$ (plateau in time).
At times of the front crossing the observation point, $\Pxt$ can be collapsed by the
rescaling of time with the distance $x$ as shown in right panel of Fig.~\ref{fig:S2}.
The above collapse gives thus evidence that the wave-front propagates diffusively, meaning $\Xf\sim t$.

 \begin{figure}[h!]
 \includegraphics[width=0.8\columnwidth]{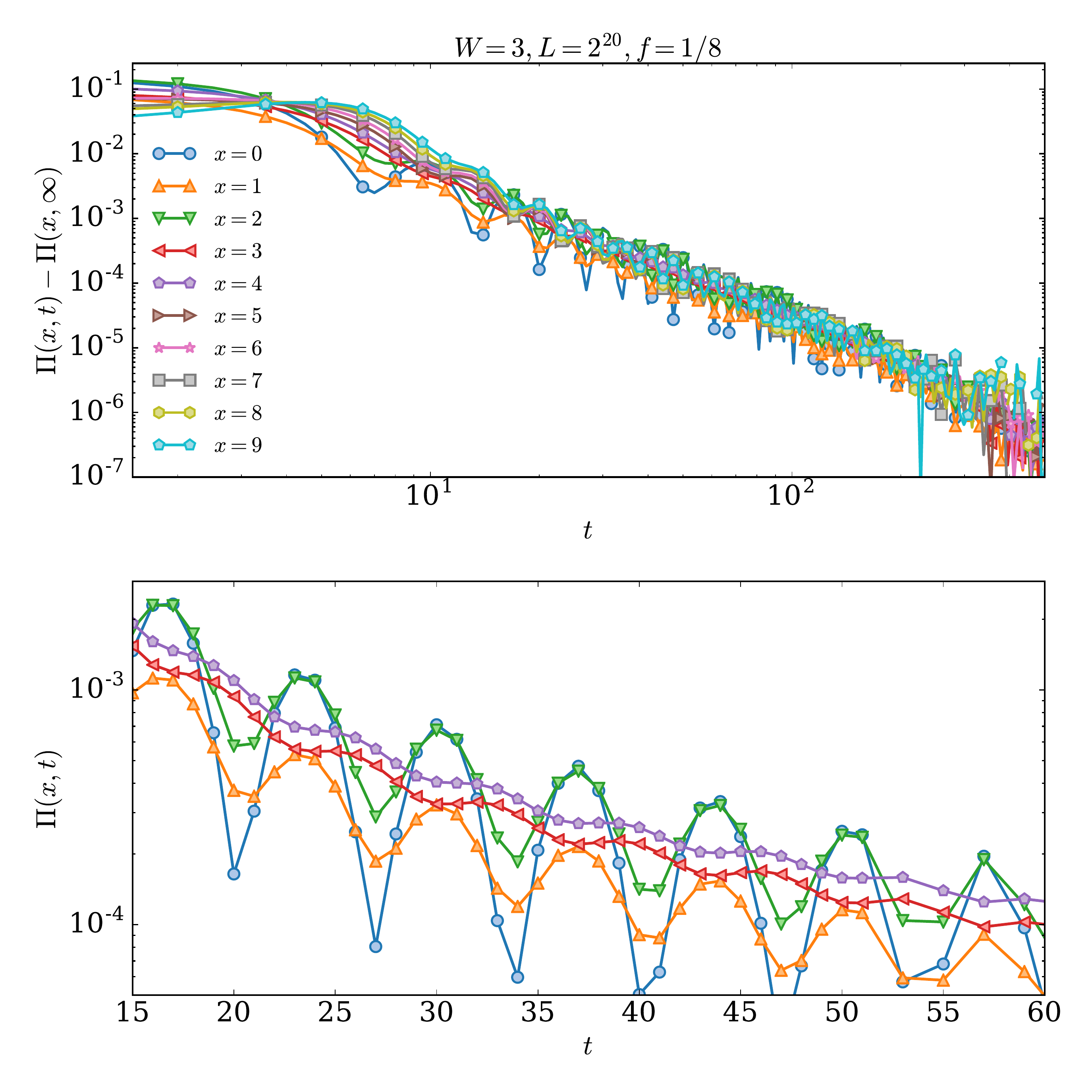}
 \caption{Left panel: $\Pxt-\Pxinf$ at $W=3$ as a function of time for several $x$. After subtracting the saturation value $\Pxinf$, for $x>\Xf$, $\Pxt \propto \Rt$ given by Eq.~\eqref{eq:return_erg}.
 Right panel: Zoom of $\Pxt$ to underline the oscillatory behavior in Eq.~\eqref{eq:return_erg}.}
\label{fig:S3}
\end{figure}
After the wave-front passes the observation point, $x<\Xf$, $\Pxt$ starts to decay in time.
This decay inside a wavepacket is governed by the return probability decay given for this case by Eq.~\eqref{eq:return_erg}.
Indeed, in the upper panel of Fig.~\ref{fig:S3} demonstrates the collapse of $\Pxt-\Pxinf$ for a wide range of distances $x$ including $x=0$
corresponding to the return probability.
The lower panel of Fig.~\ref{fig:S3} zooms the oscillations showing their in-phase behavior and $x$-independent period.

\section{Intermediate disorder}\label{SM:Sec3_intermediate_W}
This Supplemental Note is aimed to support the analysis made in the main text by showing complimentary data in the dynamically non-ergodic phase,
$0.4\lesssim W/\WAT \lesssim 0.7$.

\begin{figure}
 \includegraphics[width=1.\columnwidth]{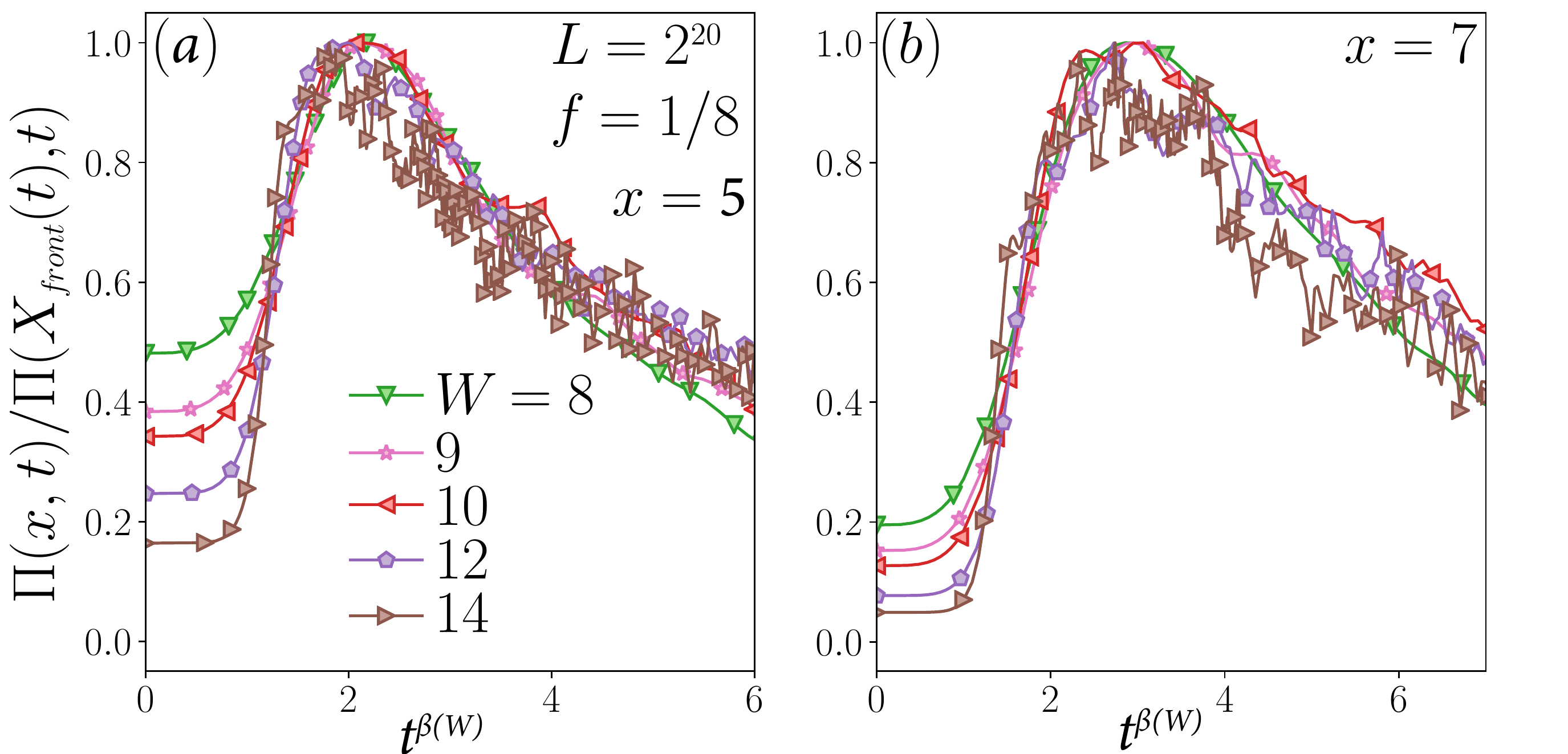}

 \caption{The figures show the collapse of the maximum in $\Pi(x,t)$ for several disorder strength $W=8,9,10,12,14$ for two $X=5,7$ respectively. The time has been rescaled as $t^{\beta(W)}$ with $\beta(W) = 1-W/W_{AT}$.}
 \label{fig:Sfig4}
 \end{figure}

We start by showing data for several disorder strengths. Figure~\ref{fig:Sfig4} 
shows $\Pi(x,t)$ for different $W$ at two different values of the distance $x$ ($x=5,7$).
The time has been properly rescaled with the exponent $\beta(W)$ to collapse the wave-fronts for several $W$,  supporting the subdiffusive dynamics $X(t)\sim t^{\beta(W)}$.

Indeed, here we focus on the regime of the return probability decaying as a stretched exponential $\Pxt\sim e^{-\Gamma t^{\beta(W)}}$, with
the exponent $\beta(W)$ approximated by $\beta(W)\approx 1-W/\WAT$.
In particular, we show the data for the disorder strength $W=12$ and the energy scale $\delta E =  f E_{\text{BW}}$ with $f=1/8$.
\begin{figure}[tb]
 \includegraphics[width=0.9\columnwidth]{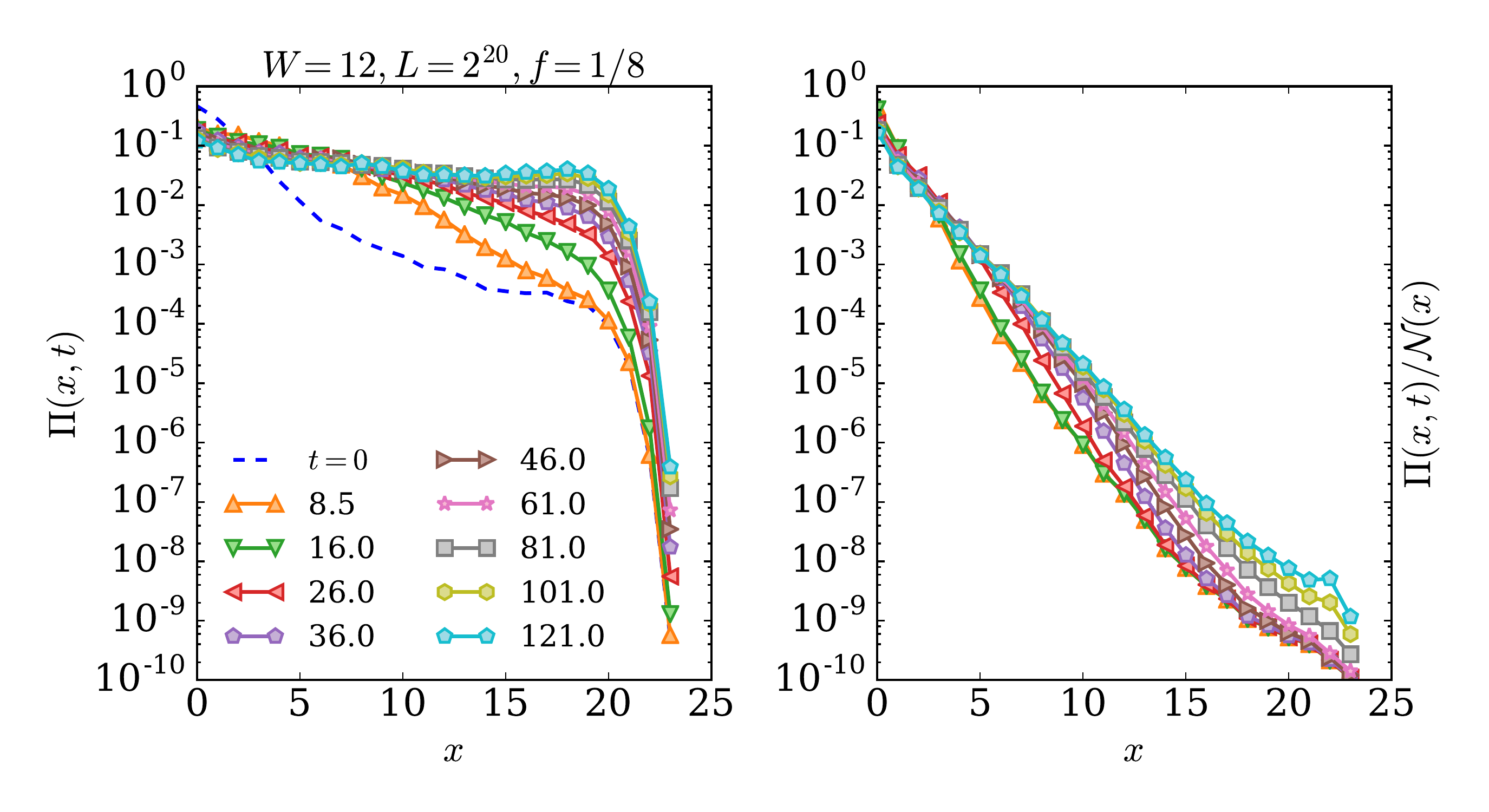}
 \caption{Left panel: $\Pxt$ at  $W=12$ as a function of the distance $x$ and several times. Right panel: $\Pxt$ renormalized
by the mean number of sites $\N(x)$  at some distance $x$ from an initial site state $|x_0 \rangle$.}
\label{fig:S4}
\end{figure}
\begin{figure}[tb]
 \includegraphics[width=0.9\columnwidth]{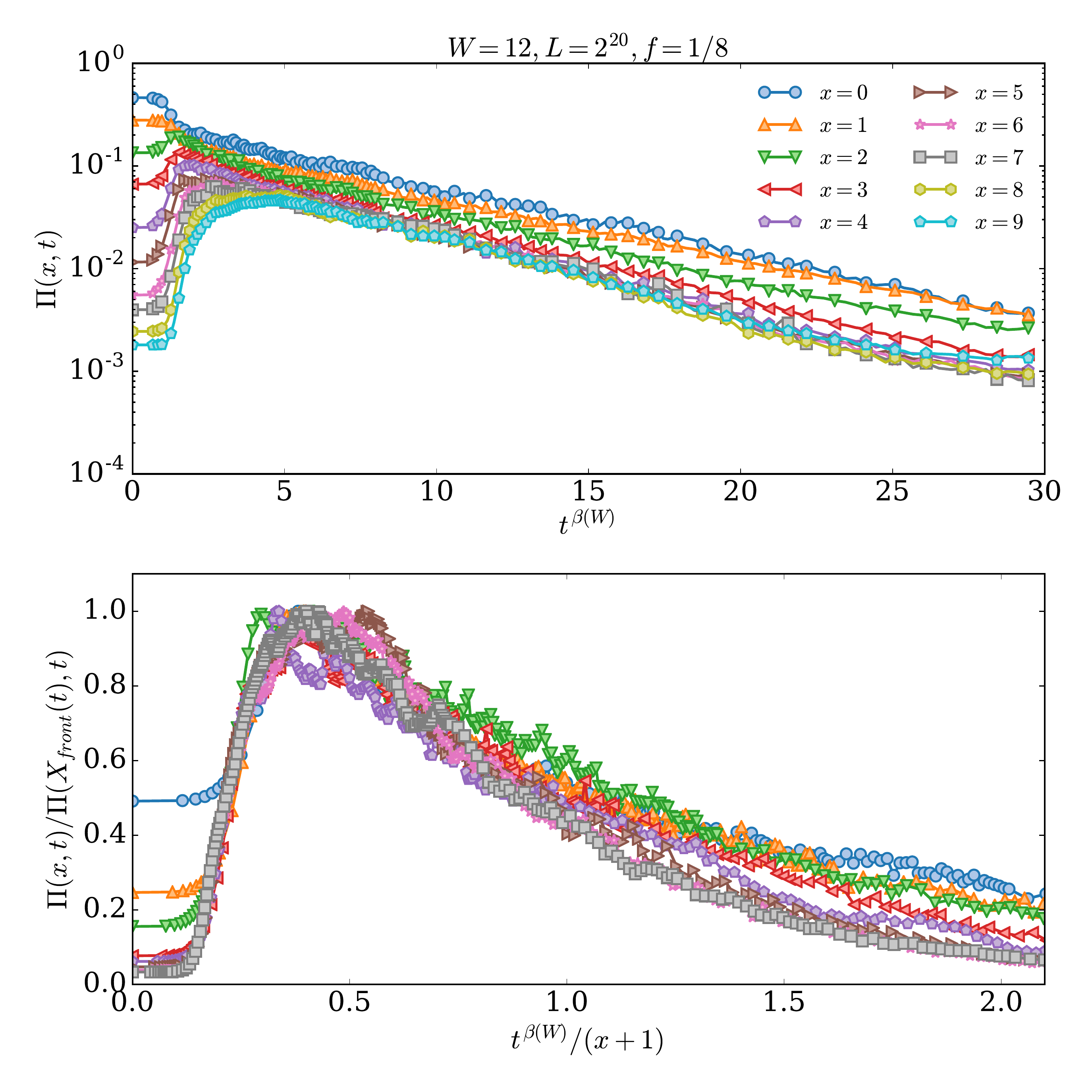}
 \caption{Left panel: $\Pxt$ at  $W=12$ as a function of time and for several $X$. The time scale has been rescaled to show $\Pxt \sim e^{-\Gamma t^{\beta(W)}}$ with $\beta(W)\approx 1-W/\WAT$.  Right panel: The collapse of the wave-front, $\Xf\sim t^{\beta(W)}$.}
\label{fig:S5}
\end{figure}
First, let's consider the short time behavior given in Fig.~\ref{fig:S4}.
At $t=0$ (dashed line in Fig~\ref{fig:S4}) the probability distribution is well localized close to the initial state,
with $1 \lesssim X(0) \lesssim 2$.
As time evolves a wave-front transfers most of the weight of $\Pxt$ to the most distant sites $x\approx D$ where $D\simeq \ln (L)/\ln K$ is the diameter of the graph.
It is clear from the plots that at this disorder strength $W=12$ compared to the one that we used in the main text
the transport slows down.
Indeed, at $W=12$ the Thouless time can be estimated to be around $\tTh\approx 10^2$, while at $W=8$ it is at least five times smaller $\tTh(W=12)/\tTh(W=8)\approx 5$.

The same universal dynamics described by the semiclassical wave-front propagation and mentioned in the main text is also observed here.
Indeed, as soon as $x>\Xf$, $\Pxt$ decays as the return probability $\Pxt\propto \Rt \sim e^{-\Gamma t^{\beta(W)}}$, with $\beta(W)$
well approximated by $1-W/\WAT$ as shown in the upper panel of Fig.~\ref{fig:S5}.
The wave-front collapse, the lower panel of Fig.~\ref{fig:S5}, is quantitatively consistent with the
the sub-diffusive propagation of the wave-front $\Xf\sim t^{\beta(W)}$ as well.

\section{Consistence of both collapses Eqs. (7) and (8) for $\Pxt$}\label{SM:Sec4_Collapse}
In this Note we show that two types of collapses used in the main text as Eqs. (7~-~8) are consistent and find the corresponding functions $f(x)$ and $g(x)$.
Indeed, in equations~(6) and~(8) of the main text it is claimed that
the probability $\Pxt$ factorizes in two intervals $x>\Xf$ and $x<\Xf$ respectively as
\be\label{SM:P(x,0)}
 \Pxt \approx \PxO \
\ee
and
\be\label{SM:Factorization}
\Pxt-\Pxinf = g(x)\lb\Rt-\Rinf\rb \ .
\ee
These equations together with the second collapse, Eq.~(7), given by
\be\label{SM:front_X_f-x}
{\Pxt-\Pxinf}= \Pi(\Xf,t) f(\Xf-x) \,
\ee
are valid only up to the diameter $x<D = \ln L/\ln K$ of the graph and should be supplied by
the normalization condition
\be\label{SM:Pxt_norm}
\sum_x \Pxt = 1 \ .
\ee

Substituting Eqs.~\eqref{SM:Factorization} and~\eqref{SM:P(x,0)} into the normalization condition~\eqref{SM:Pxt_norm} one
immediately obtains
\be\label{SM:g(x)_eq}
\delta R(X)\sum_{x=0}^{X} g(x) = \sum_{x=0}^{X}\lb \PxO - \Pxinf\rb
\ee
where we replace $\Xf$ by an arbitrary $X$ and used the notation
\be
\delta R(\Xf) = \Rt-\Rinf \ .
\ee

Equation~\eqref{SM:g(x)_eq} can be easily solved by taking the derivative of $\sum_{x=0}^{X} g(x)$ over $X$
\begin{multline}
g(x)\delta R(x) = \PxO - \Pxinf
-\\
\frac{\delta R'(x)}{\delta R(x)}
\sum_{x'=0}^{x}\lb \Pi(x',0)-\Pi(x',\infty)\rb \ .
\end{multline}
The collapse~\eqref{SM:Factorization} rewritten in terms of the function
$g(x)\delta R(x)$ at $x<\Xf$ as
\begin{multline}
\Pxt-\Pxinf= \frac{\delta R(\Xf)}{\delta R(x)} g(x)\delta R(x) \
\end{multline}
is clearly consistent with \eqref{SM:front_X_f-x} provided
the return probability written in terms of the wavefront decays exponentially
\be
\delta R(x) \sim e^{-\lambda x} \ .
\ee
This determines the function $f(x) \sim \delta R(x) \sim e^{-\lambda x}$ and
the amplitude $\Pi(\Xf,t)$ of $\Pxt$ as
\begin{multline}
\Pi(\Xf,t) = 
\Pi(\Xf,0) - \Pi(\Xf,\infty)
-\\
\frac{\delta R'(\Xf)}{\delta R(\Xf)}
\sum_{x'=0}^{\Xf}\lb \Pi(x',0)-\Pi(x',\infty)\rb \
\end{multline}
consistent with the claim in the main text.

\end{document}